\documentclass[12pt,epsf,qsymbols]{article}
\pdfoutput=1
\usepackage[utf8]{inputenc}
\usepackage[normalem]{ulem}
\usepackage[english]{babel}
\usepackage{tabularx}
\usepackage{changepage}
\usepackage{array}
\usepackage{graphics}
\usepackage[pdftex]{graphicx}
\usepackage{psfrag}
\usepackage{epsfig}
\usepackage{subfig}
\usepackage{mathtools}
\usepackage{amssymb}
\usepackage{pifont}
\usepackage{bbold}
\usepackage{setspace}
\usepackage{rotating}
\usepackage{colortbl}
\usepackage{longtable}
\usepackage{slashed}
\usepackage{braket}
\usepackage{lineno}
\makeatletter
\usepackage{textcomp}
\usepackage[usenames,dvipsnames]{xcolor}
\usepackage{relsize}
\usepackage{cite}

\usepackage{arydshln}

\usepackage{float}
\usepackage{multirow}
\usepackage{tikz}
\usetikzlibrary{decorations.pathmorphing,shapes}

\usepackage{lipsum}

\usepackage{bbm}
\usepackage{bm}
\usepackage{enumitem}
\usepackage{amsmath}
\usepackage{verbatim}

\usepackage{hyperref}
\hypersetup{colorlinks,citecolor=nicegreen,linkcolor=niceblue}
\hypersetup{colorlinks=true}

\usepackage{pifont}

\usepackage{calligra}

\usepackage{geometry}

\allowdisplaybreaks

\setlongtables

\newcommand{\bea}{\begin{eqnarray}}
\newcommand{\eea}{\end{eqnarray}}
\newcommand{\beq}{\begin{equation}}

\newcommand{\eeq}{\end{equation}}
\newcommand{\ec}{\end{center}}
\newcommand{\bc}{\begin{center}}

\newcommand{\pdir}{p\kern -5.2pt\raise 0.2ex\hbox {/}}

\newcommand{\vdir}{v\kern -5.75pt\raise 0.15ex\hbox {/}}
\newcommand{\kdir}{k\kern -5.75pt\raise 0.15ex\hbox {/}}
\newcommand{\epsdir}{\epsilon\kern -5.0pt\raise 0.15ex\hbox {/}}
\newcommand{\bvdir}{\bar{v}\kern -5.75pt\raise 0.15ex\hbox {/}}
\newcommand{\Ddir}{D\kern -7.75pt\raise 0.20ex\hbox {/}}
\newcommand{\Adir}{A\kern -7.75pt\raise 0.20ex\hbox {/}}
\newcommand{\ldir}{l\kern -5.0pt\raise 0.2ex\hbox{/}}
\newcommand{\varepsdir}{\varepsilon\kern -5.5pt\raise 0.15ex\hbox{/}}

\makeatother

\DeclareMathAlphabet{\mathcalligra}{T1}{calligra}{m}{n}
\DeclareFontShape{T1}{calligra}{m}{n}{<->s*[2.2]callig15}{}

\definecolor{lightgray}{rgb}{0.9,0.9,0.9}
\definecolor{niceblue}{rgb}{0.15,0.15,0.6}
\definecolor{nicegreen}{rgb}{0.1,0.5,0.1}
\definecolor{Red}{rgb}{1.,0.,0.}

\definecolor{Green}{rgb}{0.2,.7,0.2}

\begin{document}
\unitlength = 1mm

\thispagestyle{empty} 
\vspace*{1.8cm}
\begin{center}
\textbf{\Large Lepton flavor violation in exclusive \\[0.3em] 
$b\to d\ell_i\ell_j$ and $b\to s\ell_i\ell_j$ decay modes}
\vskip 1.6cm
\scalebox{.88}{\large \textsc{D.~Be\v{c}irevi\'c$^{\,a}$, F.~Jaffredo$^{\,b}$, J.P.~Pinheiro$^{\,c}$, O.~Sumensari$^{\,a}$}} \\
\vskip 0.9 cm
\textsl{$^a$~Universit\'e Paris-Saclay, CNRS/IN2P3, IJCLab, 91405 Orsay, France}\\
\textsl{$^b$~INFN, Sezione di Pisa, Largo Bruno Pontecorvo 3, I-56127 Pisa, Italy}\\
\textsl{$^c$~Departament de Física Quàntica i Astrofísica and Institut de Ciències del Cosmos, Universitat de Barcelona, Diagonal 647, E-08028 Barcelona, Spain}
\vskip 1.65cm
\end{center}
\vskip 0.85cm

\begin{abstract}
We discuss the exclusive lepton flavor violating (LFV) decays modes based on $b\to d\ell_i\ell_j$ and $b\to s\ell_i\ell_j$ by considering the ground state mesons and baryons. After spelling out the expressions for such decay rates in a low energy effective theory which includes generic contributions arising from physics beyond the Standard Model (BSM), we show that the experimental bounds on meson decays can be used to bound the corresponding modes involving baryons. We find, for example, $\mathcal{B}(\Lambda_b\to \Lambda\mu\tau)\lesssim 4\times 10^{-5}$. We also consider two specific models and constrain the relevant LFV couplings by using the low energy observables. In the first model we assume the Higgs mediated LFV and find the resulting decay rates to be too small to be experimentally detectable. We also 
emphasize that the regions favored by the bounds $\mathcal{B}(h\to\mu\tau)^\mathrm{Atlas}$ and  $\mathcal{B}(h\to e\tau)^\mathrm{Atlas}$ are not compatible with $\mathcal{B}(\mu\to e\gamma)^\mathrm{MEG}$ to $1\sigma$. In the second model we assume LFV mediated by a heavy $Z'$ boson and find that the  corresponding $b$-hadron branching fractions can be $\mathcal{O}(10^{-6})$, thus possibly within experimental reach at LHCb and Belle~II. 
\end{abstract}

\setcounter{page}{1}
\setcounter{footnote}{0}
\setcounter{equation}{0}
\noindent

\renewcommand{\thefootnote}{\arabic{footnote}}
 
\setcounter{footnote}{0}

\clearpage

\section{Introduction}
\label{sec:intro}
In the past decade we witnessed a huge effort in the high energy physics community to better understand the exclusive decays based on $b\to s l l$ ($l\in \{e,\mu\}$) because their detailed experimental analysis became feasible at the LHC~\cite{LHCb:2015svh,CMS:2018qih}. 
The measured angular distributions of $B\to K^{(\ast)} \mu\mu$ were compared with theoretical predictions, eventually  unveiling serious difficulties related to the treatment of hadronic uncertainties. In that situation, several apparent discrepancies between theory and experiments could not be unambiguously attributed to the presence of physics beyond the Standard Model (BSM)~\cite{Ciuchini:2021smi}. The experimental deviations from lepton flavor universality, on the other hand, were free from such uncertainties and there was a hope that the statistical significance of the observation that $R_{K^{(\ast)}}=\mathcal{B}'(B\to K^{(\ast)} \mu\mu)/\mathcal{B}'(B\to K^{(\ast)} ee) < 1$ would be further corroborated by the increased luminosity of experimental data. Recently, however, it was realized that one part of experimental uncertainties was not properly accounted for in the previous analyses and the new results turned out to be $R_{K^{(\ast)}} \simeq 1$~\cite{LHCb:2022vje}, in full agreement with the Standard Model (SM) predictions~\cite{Hiller:2003js}. In such a situation, the Lepton Flavor Violating (LFV) modes seem to be the best probes of BSM physics: (i) They are forbidden in the SM. (ii) A significant part of hadronic uncertainties present in the lepton flavor conserving (LFC) case (related to a coupling to the $c\bar c$-resonances) is absent when describing the LFV modes. (iii) These decay channels are studied experimentally, and the Belle-II and LHCb bounds on their branching fractions are not only competitive with those already established at Belle and BaBar, but they will supersede them in the years to come. There are many possible scenarios of physics BSM that could produce the effects of LFV, see Ref.~\cite{Kuno:1999jp} for a recent review. In this paper, they are treated generically through a low-energy effective field theory (EFT).
While the exclusive $b\to s l l$ decays were considered to be the main window to the particle content of physics BSM, the similar $b\to d l l$ decays received much less attention in the literature~\cite{Fuentes-Martin:2019mun}. The first reason for that is that they are Cabibbo suppressed with respect to $b\to s l l$. Furthermore, they are not only plagued by the $c\bar c$-resonances but also by the $u \bar u$ ones, making the comparison between theory and experiment ever more challenging. In the corresponding LFV modes  the problem of hadronic resonances in leptonic spectra is not present and therefore the $b\to d \ell_i \ell_j$ ($i\neq j$) decays are equally interesting in our quest for the experimental signatures of physics BSM. 

In this paper we provide the general expressions relevant to the studies of $B_{s,d}\to \ell_1 \ell_2$, $B \to P \ell_1 \ell_2$ ($P=\pi,K$), $B \to V \ell_1 \ell_2$ ($V=\rho,K^*$), and $\Lambda_b \to L \ell_1 \ell_2$ ($L=\Lambda, n$) channels. We will then consider two particular scenarios of New Physics arising from LFV couplings to: (a) the SM-like Higgs state, and (b) a heavy vector boson $Z^\prime$. We will derive the bounds on the corresponding LFV modes, which will be obtained not only by combining various low-energy constraints but also by imposing high-$p_T$ LHC constraints.

In the remainder of this paper we first, in Sec.~\ref{sec:eft}, remind the reader of the low energy effective theory and of the expressions for the leptonic and semileptonic LFV $B$-meson decays in Sec.~\ref{ssec:B-reminder}. In Sec.~\ref{sec:lambda} the general expressions for the similar baryon LFV decays are provided and the bound on the LFV modes involving baryons are derived from the bounds on similar decays involving mesons. Further phenomenology is discussed in Sec.~\ref{sec:concrete} where we consider two concrete models in which we use the low- and high-energy observables as constraints to obtain the bounds on the desired LFV branching fractions. Our findings are summarized in Sec.~\ref{sec:conclusion}.

\section{Effective theory}
\label{sec:eft}

The generic effective Hamiltonian involving dimension-six operators to describe the LFV transitions $b\to q\ell_i^- \ell_j^{+}$, with $q\in \lbrace d,s\rbrace$ and $\ell_{i,j}\in\{ e,\mu,\tau\}$, can be written as~\cite{Altmannshofer:2008dz,Becirevic:2016zri},
\begin{equation}
\label{eq:hamiltonian}
\begin{split}
  \mathcal{H}_{\mathrm{eff}} = -\frac{4
    G_F}{\sqrt{2}}V_{tb}V_{tq}^* &\sum_{a=9,10,S,P}
  \Big{[} C^{q\,ij}_a(\mu)\mathcal{O}^{q\,ij}_a(\mu) + C^{\prime\, q\,ij}_{a}(\mu) \mathcal{O}^{\prime\,q\,ij}_{a}(\mu)\Big{]}
+\mathrm{h.c.}\,,
\end{split}
\end{equation}

\noindent where $G_F$ is the Fermi constant, $C_a^{q\,ij}(\mu)$ the Wilson coefficients, and $i,j$ the lepton flavor indices. The effective operators relevant to our study are:
\begin{align}
\label{eq:C_LFV}
\begin{split}
\mathcal{O}_{9}^{q\,ij}
  &=\frac{e^2}{(4\pi)^2}(\bar{q}\gamma_\mu P_{L}
    b)(\bar{\ell}_i\gamma^\mu\ell_j)\,, \qquad\qquad\hspace*{0.4cm}
\mathcal{O}_{S}^{q\,ij} =
  \frac{e^2}{(4\pi)^2}(\bar{q} P_{R} b)(\bar{\ell}_i \ell_j)\,,\\
\mathcal{O}_{10}^{q\,ij} &=
    \frac{e^2}{(4\pi)^2}(\bar{q}\gamma_\mu P_{L}
    b)(\bar{\ell}_i\gamma^\mu\gamma^5\ell_j)\,,\qquad\qquad
\mathcal{O}_{P}^{q\,ij} =
  \frac{e^2}{(4\pi)^2}(\bar{q} P_{R} b)(\bar{\ell}_i \gamma^5 \ell_j)\,.
\end{split}
\end{align}

\noindent The chirality-flipped operators $\mathcal{O}^\prime_{a}$ are obtained from $\mathcal{O}_a$ by replacing $P_L\leftrightarrow P_R$. From now on, in order to simplify our notation, we will use $C_a \equiv C_{a}^{q\,ij}$ in the situations in which there is no confusion about the flavor indices. Notice that we do not write the tensor operators in Eq.~\eqref{eq:hamiltonian} because they can only appear at dimension-eight once the full SM gauge symmetry $SU(3)_c\times SU(2)_L\times U(1)_Y$ is imposed~\cite{Alonso:2014csa}. 

In the following, unless stated otherwise, we shall combine decays with the opposite lepton charges, i.e.,
\begin{align}
\mathcal{B}(B_{q}\to\ell_i\ell_j) &\equiv  \mathcal{B}(B_{q}\to\ell_i^+\ell_j^-)+\mathcal{B}(B_{q}\to\ell_i^-\ell_j^+)\,,
\end{align}

\noindent and, similarly, for the other decay modes. One should, however, keep in mind that in some scenarios, such as those involving a $\mathcal{O}(1~\mathrm{TeV})$ leptoquark, the branching fractions with opposite leptonic charges may in fact be different~\cite{Becirevic:2016zri,Becirevic:2016oho,Bordone:2018nbg}.

\section{$B_{q}\to\ell_i\ell_j$, $B\to P\ell_i\ell_j$ and $B\to V\ell_i\ell_j$}
\label{ssec:B-reminder}

We briefly remind the reader of the results obtained for the leptonic and semileptonic LFV decays based on the $b\to q \ell_i\ell_j$ transitions and involving mesons only~\cite{Becirevic:2016zri}. These processes will be considered in Sec.~\ref{ssec:indirect} to indirectly constrain the $b$-baryon branching fractions. Current experimental situation with the exclusive processes relevant to this paper is summarized in Table~\ref{tab:exp}. In the following, we provide the relevant theoretical expressions.

\subsection{Leptonic decays} 
The simplest LFV decays of neutral $b$-mesons are the leptonic ones,  $B_{q}\to \ell_i\ell_j$, with $q=d,s$. The relevant hadronic matrix element,
\begin{align}
\langle 0|\bar{q}\gamma^\mu\gamma_5 b|\overline{B}_{q}\rangle = i f_{B_{q}} p^\mu\,,
\end{align}

\noindent is parametrized in terms of $f_{B_{q}}$, the $B_q$-meson decay constant, accurately determined in lattice QCD. Current world average results are: $f_{B_d}=(190.0\pm 1.3)$~MeV, and $f_{B_s}=(230.3\pm 1.3)$~MeV~\cite{FlavourLatticeAveragingGroupFLAG:2021npn}. By using the effective Hamiltonian defined in Eq.~\eqref{eq:hamiltonian}, it is then straightforward to derive  the relevant branching fraction~\cite{Becirevic:2016zri}:
\begin{align}
\label{eq:B-lep}
    \mathcal{B}(B_{q} &\to \ell_i^-\ell_j^+) = \dfrac{\tau_{B_{q}}}{64 \pi^3} \dfrac{\alpha_\mathrm{em}^2 G_F^2}{m_{B_{q}}^3}  f_{B_q}^2 |V_{tb}V_{tq}^\ast|^2 \lambda^{1/2}(m_{B_s},m_{\ell_i},m_{\ell_j})\nonumber\\*
    &\times \Bigg{\lbrace} \big{[}m_{B_{q}}^2-(m_{\ell_i}+m_{\ell_j})^2\big{]}\bigg{|}\big{(}C_9^{q\,ij}-C_{9^\prime}^{q\,ij}\big{)}(m_{\ell_i}-m_{\ell_j})+ \big{(}C_S^{q\,ij}-C_{S^\prime}^{q\,ij}\big{)}\,\dfrac{m_{B_{q}}^2}{m_b+m_q}\bigg{|}^2 \\
    &\quad+ \big{[}m_{B_{q}}^2-(m_{\ell_i}-m_{\ell_j})^2\big{]}\bigg{|}\big{(}C_{10}^{q\,ij}-C_{{10}^\prime}^{q\,ij}\big{)}(m_{\ell_i}+m_{\ell_j})+ \big{(}C_P^{q\,ij}-C_{P^\prime}^{q\,ij}\big{)}\,\dfrac{m_{B_{q}}^2}{m_b+m_q}\bigg{|}^2 \Bigg{\rbrace}\,,\nonumber
\end{align}

\noindent where $\lambda(a,b,c)\equiv[a^2-(b-c)^2][a^2-(b+c)^2]$. Notice, in particular, that the analogous expression for the decay with opposite lepton charges can be obtained by flipping the sign in front of the term proportional to $C_{9}-C_{9^{\prime}}$ in the above expression. In the $B_s$ system, one should also account for the effects of $B_s$--$\overline{B_s}$ oscillation in the above expression since the time-dependence of the decays rates are integrated out in experiment, see Ref.~\cite{DeBruyn:2012wj}.

\begin{table}[!t]
\renewcommand{\arraystretch}{1.65}
\centering
\resizebox{\columnwidth}{!}{
\begin{tabular}{|c|cc||c|cc||c|cc|}
\hline
Decay & Exp.~Limit & Ref. & Decay & Exp.~Limit & Ref. & Decay & Exp.~Limit & Ref. \\ \hline\hline
$B^0 \to e \mu$ & $1.0\times 10^{-9}$ & \cite{LHCb:2017hag} & $B^0 \to e \tau$ & $1.6\times 10^{-5}$ & \cite{Belle:2021rod} & $B^0 \to \mu \tau$ & $1.2\times 10^{-5}$ & \cite{LHCb:2019ujz} \\
$B^+ \to \pi^+ e \mu$ & $9.2\times 10^{-8}$ & \cite{BaBar:2007xeb} & $B^+ \to \pi^+ e \tau$ & $7.5\times 10^{-5}$ & \cite{BaBar:2012azg} & $B^+ \to \pi^+ \mu \tau$ & $7.2\times 10^{-5}$ & \cite{BaBar:2012azg} \\  \hline
$B_s \to e \mu$ & $5.4\times 10^{-9}$ & \cite{LHCb:2017hag} & $B_s \to e \tau$ & $1.4\times 10^{-3}$ & \cite{Belle:2023jwr} & $B_s \to \mu \tau$ & $3.4\times 10^{-5}$ & \cite{Belle:2021rod} \\
$B^+ \to K^+ e \mu$ & $1.8\times 10^{-8}$ & \cite{LHCb:2019bix} & $B^+ \to K^+ e \tau$ & $3.1\times 10^{-5}$ & \cite{Belle:2022pcr} & $B^+ \to K^+ \mu \tau$ & $3.1\times 10^{-5}$ & \cite{Belle:2022pcr} \\
$B^0 \to K^{\ast 0} e \mu$ & $1.0\times 10^{-8}$ & \cite{LHCb:2022lrd} & $B^0 \to K^{\ast 0} e \tau$ & -- & & $B^0 \to K^{\ast 0} \mu \tau$ & $1.8\times 10^{-5}$ & \cite{LHCb:2022wrs} \\ 
$B_s \to \phi e \mu$ & $1.6\times 10^{-8}$ & \cite{LHCb:2022lrd} & $B_s \to \phi e \tau$ & -- &  & $B_s \to \phi \mu \tau$ & $2.0\times 10^{-5}$ & \cite{LHCb:2024wve} \\
\hline

\end{tabular}}
\caption{\small \sl Experimental limits on (semi)leptonic LFV decay of $B$-mesons at 90\%~CL. We only list the upper limits on decays with both combinations of lepton charges, i.e.,~$\ell_i \ell_j \equiv \ell_i^+ \ell_j^- +  \ell_i^-\ell_j^+$.}
\label{tab:exp}
\end{table}

\subsection{Semileptonic modes} 
The decays $B\to P \ell_i\ell_j$, with the pseudoscalar meson in the final state $P=\pi$ or $K$, and $B\to V \ell_i\ell_j$ with the vector meson $V=\rho$, $K^\ast$ or $\phi$, have been extensively studied experimentally, cf.~Table~\ref{tab:exp}. The relevant explicit expressions for their differential branching fractions are given in Ref.~\cite{Becirevic:2016zri} and also in Refs~\cite{Gratrex:2015hna,Plakias:2023esq}. After integrating over the phase space, it is possible to write the branching fractions as 
\begin{align}
\label{eq:B-semilep}
    \mathcal{B}(B\to M \ell_i^- \ell_j^+) &\times 10^9  = a_9\,|C_9+C_{9^\prime}|^2 +a_{10}\,|C_{10}+C_{{10}^\prime}|^2 +a_S\,|C_S+C_{S^\prime}|^2 +a_{P}\,|C_{P}+C_{{P}^\prime}|^2 \nonumber \\[0.35em] 
    &+b_9\,|C_9-C_{9^\prime}|^2 +b_{10}|C_{10}-C_{{10}^\prime}|^2 +b_S|C_S-C_{S^\prime}|^2 +b_{P}|C_{P}-C_{{P}^\prime}|^2 \\[0.35em]\nonumber
    &+c_{9S}\,\mathrm{Re}[(C_9\pm C_{9^\prime})(C_S\pm C_{S^\prime})^\ast] +c_{10P} \,\mathrm{Re}[(C_{10}\pm C_{{10}^\prime})(C_P\pm C_{P^\prime})^\ast] \,,
\end{align}
\noindent where the upper (lower) sign corresponds to $M=P$ ($V$), and $a_I$, $b_I$, $c_I$ are the coefficients, with $I\in \lbrace 9,10,S,P,9S,10P\rbrace$, which are obtained upon integration of the kinematical factors and the form factors, often referred to as the {\sl magic numbers}. For simplicity, we omit the flavor indices. For $B\to P$ decays, the coefficients $b_{9,10}$ and $b_{S,P}$ vanish since the axial and pseudoscalar matrix elements are identically zero. For $P\to V$ decays, instead, the coefficients $a_{S,P}$ vanish. Moreover, in the limit in which one of the lepton masses is negligible, say $m_{\ell_i}\ll m_{\ell_j}$, one can show that~\cite{Becirevic:2016zri,Plakias:2023esq}:
\begin{align}\label{eq:massless}
a_9 &\simeq a_{10}\,,\qquad a_S\simeq a_P\,, \qquad c_{9S} \simeq -c_{10P}\,, \qquad
b_9 \simeq b_{10}\,,\qquad b_S\simeq b_P\,, 
\end{align}
valid up to corrections $\mathcal{O}(m_{\ell_i}/m_{\ell_j})$.

\begin{table}[!p]
\renewcommand{\arraystretch}{1.7}
\centering
\resizebox{\columnwidth}{!}{
\begin{tabular}{|c|cccccccccc|}
\hline 
Decay  & $a_9$ & $b_{9}$ & $a_{10}$ & $b_{10}$ & $a_S$ & $b_S$ & $a_P$ & $b_P$ & $c_{9S}$ & $c_{10P}$ \\ \hline\hline
$B^+\to \pi^+ e^- \mu^+$ & $0.65(5)$& $0$ & $0.65(5)$& $0$ &$1.05(7)$ & $0$ & $1.05(7)$ & $0$ & $-0.069(5)$ & 0.069(5) \\
$B^+\to \pi^+ e^- \tau^+$ &$0.45(3)$ & 0 & $0.45(3)$& $0$ & 0.66(4)& $0$ & 0.66(4) & $0$ & $-0.57(4)$ & 0.57(4) \\
$B^+\to \pi^+ \mu^- \tau^+$ & $0.44(3)$& 0 & $0.45(3)$& $0$ & 0.64(4)& $0$ &0.68    (4)  & $0$ & $-0.52(3)$ &0.63(4)  \\ \hline
$B_s\to \bar{K^0} e^- \mu^+$ & $0.36(8)$& $0$ & $0.36(8)$& $0$ &$0.60(8)$ & $0$ & $0.60(8)$ & $0$ & $-0.044(7)$ & 0.044(7) \\
$B_s\to \bar{K^0} e^- \tau^+$ &$0.31(8)$ & 0 & $0.31(8)$& $0$ & 0.38(4)& $0$ & 0.38(4) & $0$ & $-0.33(4)$ & 0.33(4) \\
$B_s\to \bar{K^0} \mu^- \tau^+$ & $0.27(4)$& 0 & $0.28(4)$& $0$ & 0.36(4)& $0$ &0.39(4)  & $0$ & $-0.30(4)$ &0.37(5)  \\ \hline
$B_c^+\to D^+ e^- \mu^+$ & $0.12(3)$& $0$ & $0.12(3)$& $0$ &$0.15(2)$ & $0$ & $0.15(2)$ & $0$ & $-0.012(2)$ & 0.012(2) \\
$B_c^+\to D^+ e^- \tau^+$ &$0.08(1)$ & 0 & $0.08(1)$& $0$ & 0.08(1)& $0$ & 0.08(1) & $0$ & $-0.09(1)$ & 0.09(1) \\
$B_c^+\to D^+ \mu^- \tau^+$ & $0.08(1)$& 0 & $0.09(1)$& $0$ & 0.08(1)& $0$ &0.09(1)  & $0$ & $-0.10(1)$ &0.08(1)  \\ \hline
\hline
$B^+\to K^+ e^- \mu^+$ &18.7(7) & $0$ &18.7(7) & 0 &26.3(7) & $0$ & 26.3(7) & $0$ & $-2.05(7)$ & 2.05(7) \\
$B^+\to K^+ e^- \tau^+$ & 11.7(3)& $0$ & 11.7(3)& $0$ & 15.0(4)& $0$ & 15.0(4) & $0$ & $-15.1(4)$ & 15.1(4) \\
$B^+\to K^+ \mu^- \tau^+$ &11.4(3) & $0$ &11.8(4) & $0$ & 14.4(4)& $0$ & 15.5(4) & $0$ & $-13.4(4)$ &  16.5(5)  \\ \hline
$B_c^+\to D_s^+ e^- \mu^+$ &3.2(3) & $0$ &3.2(3) & 0 &5.0(2) & $0$ & 5.0(2) & $0$ & $-0.42(2)$ & 0.42(2) \\
$B_c^+\to D_s^+ e^- \tau^+$ & 2.2(2)& $0$ & 2.2(2)& $0$ & 2.7(1)& $0$ & 2.7(1) & $0$ & $-3.0(1)$ & 3.0(1) \\
$B_c^+\to D_s^+ \mu^- \tau^+$ &2.1(2) & $0$ &2.2(2) & $0$ & 2.6(1)& $0$ & 2.8(1) & $0$ & $-2.6(1)$ &  3.3(1)  \\ \hline
\end{tabular}}
\caption{\small \sl Numerical coefficients $a_I \equiv a_I^{q\, ij}$ defined in Eq.~\eqref{eq:B-semilep} for the most relevant LFV pseudoscalar meson decays based on the transitions $b\to d\ell_i^-\ell_j^+$ and $b\to s\ell_i^-\ell_j^+$. The coefficients for the channels with the opposite lepton charges in the final state are identical, with the exception of $c_{9S}\to -c_{9S}$. Note that the coefficients are multiplied by $10^{-9}$, cf.~Eq.~\eqref{eq:B-semilep}. }
\label{tab:magic-numbers-mesons} 
\end{table}

\begin{table}[!p]
\renewcommand{\arraystretch}{1.7}
\centering
\resizebox{\columnwidth}{!}{
\begin{tabular}{|c|cccccccccc|}
\hline 
Decay  & $a_9$ & $b_{9}$ & $a_{10}$ & $b_{10}$ & $a_S$ & $b_S$ & $a_P$ & $b_P$ & $c_{9S}$ & $c_{10P}$ \\ \hline\hline
$B^0\to \rho^0 e^- \mu^+$ & $0.27(4)$ & $1.2(3)$ & $0.27(4)$ & $1.2(3)$ & $0$ & $0.48(10)$ & $0$ & $0.48(10)$ & $-0.05(2)$ & $0.05(2)$\\
$B^0\to \rho^0 e^- \tau^+$ & $0.14(2)$ & $0.66(14)$ & $0.14(2)$ & $0.66(14)$ & $0$ & $0.22(5)$ & $0$ & $0.22(5)$ & $-0.28(5)$ & $.28(5)$\\
$B^0\to \rho^0 \mu^- \tau^+$ & $0.16(2)$ & $0.73(14)$ & $0.14(2)$ & $0.69(13)$ & $0$ & $0.22(4)$ & $0$ & $0.25(5)$ & $-0.26(6)$ & $0.33(6)$\\ 
\hline

\hline
$B^0\to \omega e^- \mu^+$ & $0.23(4)$ & $1.1(3)$ & $0.23(4)$ & $1.1(3)$ & $0$ & $0.41(9)$ & $0$ & $0.41(9)$ & $-0.05(1)$ & $0.05(1)$\\
$B^0\to \omega e^- \tau^+$ & $0.12(2)$ & $0.60(13)$ & $0.12(2)$ & $0.60(13)$ & $0$ & $0.19(4)$ & $0$ & $0.19(4)$ & $-0.24(5)$ & $0.24(5)$\\
$B^0\to \omega \mu^- \tau^+$ & $0.12(2)$ & $0.61(13)$ & $0.11(2)$ & $0.58(13)$ & $0$ & $0.18(4)$ & $0$ & $0.20(4)$ & $-0.21(4)$& $0.27(6)$\\ 
\hline

$B_s\to K^{\ast 0} e^- \mu^+$ & $0.20(2)$ & $1.2(1)$& $0.20(2)$ & $1.2(1)$ & $0$ & $0.49(4)$ & $0$ & $0.49(4)$ & $-0.05(1)$ & $0.05(1)$ \\
$B_s\to K^{\ast 0} e^- \tau^+$ & $0.11(1)$ & $0.65(6)$ & $0.11(1)$ & $0.65(6)$ & $0$ & $0.23(2)$ & $0$& $0.23(2)$ & $-0.29(2)$ & $0.29(2)$\\
$B_s\to K^{\ast 0} \mu^- \tau^+$ & $0.12(1)$  & $0.66(6)$ & $0.10(1)$ & $0.63(6)$ & $0$ & $0.22(2)$ & $0$ & $0.24(2)$ & $-0.25(2)$ & $0.32(2)$\\ 

\hline\hline

\hline
$B^0\to K^{\ast 0} e^- \mu^+$ & $6.3(6)$ & $31.5(3)$ & $6.3(7)$ & $31.5(3)$ & $0$ & $9.8(10)$ & $0$ & $9.8(10)$ & $-1.2(3)$ & $1.2(3)$ \\
$B^0\to K^{\ast 0} e^- \tau^+$ & $3.1(3)$ & $16.6(17)$ & $3.1(3)$ & $16.6(17)$ & $0$ & $4.3(4)$ & $0$ & $4.3(4)$ & $-5.8(4)$ & $5.8(4)$ \\
$B^0\to K^{\ast 0} \mu^- \tau^+$ & $3.3(3)$ & $17.0(17)$ & $2.9(2)$ & $16.0(16)$ & $0$ & $4.0(4)$ & $0$ & $4.4(4)$ & $-5.0(5)$& $6.4(5)$\\ 

\hline

$B_s\to\phi\, e^- \mu^+$ & $5.5(3)$ & $34(3)$ & $5.5(3)$ & $34(3)$ & $0$ & $11(1)$ & $0$ & $11(1)$ & $-1.4(1)$ & $1.4
(1)$ \\
$B_s\to \phi \,e^- \tau^+$ & $2.8(2)$ & $17(1)$ & $2.8(2)$ & $17(1)$ & $0$ & $4.6(3)$ &$0$ & $4.6(3)$ & $-6.6(4)$ & $6.6(4)$ \\
$B_s\to \phi \,\mu^- \tau^+$ &$2.9(2)$ &$17(1)$ &$2.6(2)$ & $16(1)$ & $0$ & $4.5(3)$ & $0$ & $4.9(3)$ & $-5.6(4)$& $7.3(5)$\\ \hline

\end{tabular}}
\caption{\small \sl Numerical coefficients $a_I \equiv a_I^{q\, ij}$ defined in Eq.~\eqref{eq:B-semilep} for the most relevant LFV vector meson decays based on the transitions $b\to d\ell_i^-\ell_j^+$ and $b\to s\ell_i^-\ell_j^+$. The coefficients for the channels with the opposite lepton charges in the final state are identical, with the exception of $c_{9S}\to -c_{9S}$. Note that the coefficients are multiplied by $10^{-9}$, cf.~Eq.~\eqref{eq:B-semilep}. }
\label{tab:magic-numbers-mesons2} 
\end{table}

We computed the magic numbers appearing in Eq.~\eqref{eq:B-semilep} for a number of channels and the results are collected in Tables~\ref{tab:magic-numbers-mesons} and \ref{tab:magic-numbers-mesons2}, which is also an extended update of the results given in  Ref.~\cite{Becirevic:2016zri}. Concerning all of the $B\to P$ modes, the hadronic form factors have already been computed in lattice QCD. More specifically we use: (i) For the $B\to \pi$ and $B_s\to K$ transitions the respective results from Refs.~\cite{FermilabLattice:2015mwy,Flynn:2015mha,Colquhoun:2022atw} and Refs.~\cite{Flynn:2015mha,Bouchard:2014ypa,FermilabLattice:2019ikx} were combined in Ref.~\cite{FlavourLatticeAveragingGroupFLAG:2021npn}. (ii)  The $B\to K$ form factors computed in Refs.~\cite{Bailey:2015dka,Parrott:2022rgu}, were combined in~\cite{Becirevic:2023aov}. (iii) The form factor results relevant to $B_c\to D_{(s)} \ell_i\ell_j$ decays were presented in  Ref.~\cite{Cooper:2021bkt}. The situation with $B\to V$ decays is much less well studied on the lattice and in this paper we rely on the model calculations based on light cone QCD sum rules~\cite{Bharucha:2015bzk,Gubernari:2018wyi}.

\section{$\Lambda_b\to \Lambda \ell_i\ell_j$ and $\Lambda_b\to n \ell_i\ell_j$ decays }
\label{sec:lambda}

We now turn to the corresponding baryon decays, see also Ref.~\cite{Bordone:2021usz,Angelescu:2021lln}. 
As far as the $b\to s\ell_i\ell_j$ modes are concerned, we focus on $\Lambda_b\to \Lambda \ell_i\ell_j$ but the expressions are {\sl mutatis mutandis} applicable to the similar $(1/2)^+ \to (1/2)^+$ modes, such as $\Sigma_b^\pm \to \Sigma^\pm \ell_i\ell_j$, $\Xi_b^{0,-}\to \Xi^{0,-}\ell_i\ell_j$, for which unfortunately there are no corresponding form factors. As for the $b\to d\ell_i\ell_j$ transitions we provide the expressions for $\Lambda_b\to n \ell_i\ell_j$ but the formulas are, after obvious replacements, applicable to $\Sigma_b^+ \to p \ell_i\ell_j$, $\Xi_b^{0}\to \Lambda\ell_i\ell_j$, $\Xi_b^{-}\to \Sigma^{-}\ell_i\ell_j$. Of all these latter decays only for $\Xi_b^{0}\to \Lambda\ell_i\ell_j$ the hadronic form factors were computed~\cite{Faustov:2018ahb}, which is why we will present the numerical results for this mode too. On a more pragmatic side, the decay $\Xi_b^{0}\to \Lambda (\to p\pi)\ell_i\ell_j$ could be experimentally easier to look for than the one with neutron in the final state.

\subsection{Hadronic matrix elements}
\label{ssec:baryon-ffs}

To get a formula similar to Eq.~\eqref{eq:B-semilep}, but for the baryonic modes, we again start from the effective theory~\eqref{eq:hamiltonian}. The hadronic matrix elements are again expressed in terms of hadronic form factors. We adopt the following decomposition:
\begin{align}
\begin{split}
\label{eq:FF-baryon-1}
\langle \Lambda(k,s_\Lambda)| \bar{s}\gamma^\mu b |\Lambda_b(p,s_{\Lambda_b}) \rangle =  &\bar{u}_\Lambda(k,s_\Lambda) \Bigg[ f_0(q^2)(m_{\Lambda_b}-m_{\Lambda})\frac{q^\mu}{q^2}
 \\ 
&+f_+(q^2)\frac{(m_{\Lambda_b}+m_{\Lambda})}{\beta_+^\Lambda}\Bigg( p^\mu + k^\mu   - (m_{\Lambda_b}^2-m_{\Lambda}^2)\frac{q^\mu}{q^2} \Bigg) \\ 
&+f_\perp (q^2)\Bigg( \gamma^\mu -\frac{2 m_\Lambda}{\beta_+^\Lambda} p^\mu   - \frac{2 m_{\Lambda_b}}{\beta_+^\Lambda} k^\mu \Bigg)\Bigg] u_{\Lambda_b}(p,s_{\Lambda_b})\,,
\end{split}\\[+3mm]
\begin{split}
\label{eq:FF-baryon-2}
\langle \Lambda(k,s_\Lambda)| \bar{s}\gamma^\mu \gamma_5 b 
|\Lambda_b(p,s_{\Lambda_b}) \rangle =  & \bar{u}_\Lambda(k,s_\Lambda) \gamma_5 \Bigg[ g_0(q^2)(m_{\Lambda_b}+m_{\Lambda})\frac{q^\mu}{q^2}
 \\ 
&+g_+(q^2)\frac{(m_{\Lambda_b}-m_{\Lambda})}{\beta_-^\Lambda}\Bigg( p^\mu + k^\mu   - (m_{\Lambda_b}^2-m_{\Lambda}^2)\frac{q^\mu}{q^2} \Bigg)  \\ 
&+g_\perp (q^2)\Bigg( \gamma^\mu +\frac{2 m_\Lambda}{\beta_-^\Lambda} p^\mu   - \frac{2 m_{\Lambda_b}}{\beta_-^\Lambda} k^\mu \Bigg)\Bigg] u_{\Lambda_b}(p,s_{\Lambda_b})\, \,,
\end{split}
\end{align}
where, for definiteness, we consider the $\Lambda_b\to \Lambda$ transition,  $s_{\Lambda_{(b)}}$ is the spin of $\Lambda_{(b)}$, $q=p-k$, and $\beta^\Lambda_{\pm} = (m_{\Lambda_b}\pm m_{\Lambda})^2 - q^2$. The vector ($f_{0,+,\perp}$) and the axial ($g_{0,+,\perp}$) form factors are subject to the following constraints: (a) At $q^2=0$, $f_0(0)=f_+(0)$ and $g_0(0)=g_+(0)$. (b) At $q^2_\mathrm{max}=(m_{\Lambda_b}-m_\Lambda)^2$, $g_+(q^2_\mathrm{max})=g_\perp(q^2_\mathrm{max})$. The hadronic matrix elements of the scalar and pseudoscalar operators are easily obtained from Eqs.~\eqref{eq:FF-baryon-1}--\eqref{eq:FF-baryon-2} by virtue of the Ward identities:
\begin{align}
\label{eq:FF-baryon-3}
\langle \Lambda(k,s_\Lambda)| \bar{s} b |\Lambda_b(p,s_{\Lambda_b}) \rangle &= f_0(q^2)\dfrac{m_{\Lambda_b}-m_\Lambda}{m_b-m_s} \bar{u}_\Lambda(k,s_\Lambda) u_{\Lambda_b}(p,s_{\Lambda_b})\,,\\[0.4em]
\label{eq:FF-baryon-4}
\langle \Lambda(k,s_\Lambda)| \bar{s} \gamma_5 b |\Lambda_b(p,s_{\Lambda_b}) \rangle &= g_0(q^2)\dfrac{m_{\Lambda_b}+m_\Lambda}{m_b+m_s} \bar{u}_\Lambda(k,s_\Lambda)\gamma_5 u_{\Lambda_b}(p,s_{\Lambda_b})\,.
\end{align}
Note that in the literature an alternative decomposition of hadronic matrix elements is often adopted. Such is the case in Ref.~\cite{Faustov:2018ahb} which we use for $\Xi_b^{0}\to \Lambda\ell_i\ell_j$ and for that reason we provide the relation to the form factors used herein Appendix~\ref{app:ffs}.

\subsection{Differential decay rate}
\label{ssec:baryon-expressions}

By using the above decomposition of hadronic matrix elements, we can now express the decay rate of $\Lambda_b \to \Lambda\ell_i\ell_j$ in terms of the New Physics couplings, the hadronic form factors and the kinematical quantities, namely,
\begin{align}
\label{eq:Lb-semilep}
    &\dfrac{\mathrm{d}\mathcal{B}(\Lambda_b\to \Lambda \ell_i^- \ell_j^+)}{\mathrm{d}q^2} = \dfrac{\tau_{\Lambda_b}}{1024 \pi^5} \dfrac{\alpha_\mathrm{em}^2 G_F^2}{m_{\Lambda_b}^3} |V_{tb}V_{ts}^\ast|^2 \frac{\lambda^{1/2}_\ell\lambda^{1/2}_\Lambda}{q^2}\times \nonumber\\[0.3em]
   &\qquad \times \Bigg{\lbrace}\beta^\ell_-\beta^\Lambda_-\Bigg{[}\bigg{|}
    g_0(q^2)\frac{m_{\Lambda_b}+m_{\Lambda}}{m_b+m_s}(C_{{P}}-C_{{P}^\prime})-g_0(q^2)\frac{m_{\ell_i}+m_{\ell_j}}{q^2}(m_{\Lambda_b}+m_{\Lambda})(C_{{10}}-C_{{10}^\prime})
    \bigg{|}^2\nonumber\\
    &\qquad\qquad\qquad+\frac{1}{3}\left(2+\frac{(m_{\ell_i}+m_{\ell_j})^2}{q^2}\right)\left(f_+(q^2)^2\frac{(m_{\Lambda_b}+m_{\Lambda})^2}{q^2}+2f_\perp(q^2)^2\right)\big{|}C_{{9}}+C_{{9}^\prime}\big{|}^2\Bigg{]}
    \nonumber\\
    &\qquad\quad+ \beta^\ell_-\beta^\Lambda_+\Bigg{[}\bigg{|}
    f_0(q^2)\frac{m_{\Lambda_b}-m_{\Lambda}}{m_b-m_s}(C_{{P}}+C_{{P}^\prime})+f_0(q^2)\frac{m_{\ell_i}+m_{\ell_j}}{q^2}(m_{\Lambda_b}-m_{\Lambda})(C_{{10}}+C_{{10}^\prime})
    \bigg{|}^2\nonumber\\
    &\qquad\qquad\qquad +\frac{1}{3}\left(2+\frac{(m_{\ell_i}+m_{\ell_j})^2}{q^2}\right)\left(g_+(q^2)^2\frac{(m_{\Lambda_b}-m_{\Lambda})^2}{q^2}+2g_\perp(q^2)^2\right)\big{|}C_{{9}}-C_{{9}^\prime}\big{|}^2\Bigg{]}
    \nonumber\\
    &\qquad\quad +\beta^\ell_+\beta^\Lambda_-\Bigg{[}\bigg{|}
    g_0(q^2)\frac{m_{\Lambda_b}+m_{\Lambda}}{m_b+m_s}(C_{{S}}-C_{{S}^\prime})-g_0(q^2)\frac{m_{\ell_i}-m_{\ell_j}}{q^2}(m_{\Lambda_b}+m_{\Lambda})(C_{{9}}+C_{{9}^\prime})
    \bigg{|}^2\nonumber\\
    &\qquad\qquad\qquad+\frac{1}{3}\left(2+\frac{(m_{\ell_i}-m_{\ell_j})^2}{q^2}\right)\left(f_+(q^2)^2\frac{(m_{\Lambda_b}+m_{\Lambda})^2}{q^2}+2f_\perp(q^2)^2\right)\big{|}C_{{10}}+C_{{10}^\prime}\big{|}^2\Bigg{]}
    \nonumber\\
    &\qquad\quad+\beta^\ell_+\beta^\Lambda_+\Bigg{[}\bigg{|}
    f_0(q^2)\frac{m_{\Lambda_b}-m_{\Lambda}}{m_b-m_s}(C_{{S}}+C_{{S}^\prime})-f_0(q^2)\frac{m_{\ell_i}-m_{\ell_j}}{q^2}(m_{\Lambda_b}-m_{\Lambda})(C_{{9}}+C_{{9}^\prime})
    \bigg{|}^2\nonumber\\
    &\qquad\qquad\qquad+\frac{1}{3}\left(2+\frac{(m_{\ell_i}-m_{\ell_j})^2}{q^2}\right)\left(g_+(q^2)^2\frac{(m_{\Lambda_b}-m_{\Lambda})^2}{q^2}+2g_\perp(q^2)^2\right)\big{|}C_{{10}}-C_{{10}^\prime}\big{|}^2\Bigg{]}
   \Bigg{\rbrace}\,,
\end{align}
where 
$\beta^\ell_\pm=q^2-(m_{\ell_i}\pm m_{\ell_j})^2$, $\beta^\Lambda_\pm=(m_{\Lambda_b}\pm m_{\Lambda})^2-q^2$, $\lambda_\Lambda\equiv \lambda(\sqrt{q^2}, m_{\Lambda_b},m_{\Lambda}) = \beta^\Lambda_+\beta^\Lambda_-$ and $\lambda_\ell \equiv \lambda(\sqrt{q^2}, m_{\ell_i},m_{\ell_j}) = \beta^\ell_+\beta^\ell_-$. The above expression agrees with the one presented in Ref.~\cite{Bordone:2021usz}. Coefficients in the angular distribution of this decay are collected in Appendix~\ref{app:angular}. Since the form factors for this decay have been computed in lattice QCD for both $b\to s$ and $b\to d$ transitions, we will use the results of Refs.~\cite{Detmold:2015aaa,Detmold:2016pkz} and integrate the differential decay rate to express the total rate in a form analogous to Eq.~\eqref{eq:B-semilep}, namely, 
\begin{align}
\label{eq:B-baryon}
    \mathcal{B}(\Lambda_b \to \Lambda \ell_i^- \ell_j^+) &\times 10^9  = a_9\,|C_9+C_{9^\prime}|^2 +a_{10}\,|C_{10}+C_{{10}^\prime}|^2 +a_S\,|C_S+C_{S^\prime}|^2 +a_{P}\,|C_{P}+C_{{P}^\prime}|^2 \nonumber \\*[0.35em] 
    &+b_9\,|C_9-C_{9^\prime}|^2 +b_{10}|C_{10}-C_{{10}^\prime}|^2 +b_S\,|C_S-C_{S^\prime}|^2 +b_{P}|C_{P}-C_{{P}^\prime}|^2 \\*[0.35em]\nonumber
    &+c_{9S}\,\mathrm{Re}[(C_9+C_{9^\prime})(C_S+C_{S^\prime})^\ast] +c_{10P} \,\mathrm{Re}[(C_{10}+C_{{10}^\prime})(C_P+C_{P^\prime})^\ast] \,,
\end{align}

\noindent where $a_I \equiv a_I^{q\, ij}$, $b_I \equiv b_I^{q\, ij}$ and $c_I \equiv c_I^{q\, ij}$ are the numerical coefficients ({\em magic numbers}) that we computed and collected in Table~\ref{tab:magic-numbers-baryons}. After inspection we find that these coefficients, in the limit in which one of the outgoing leptons is massless, satisfy the same relations as those given in Eq.~\eqref{eq:massless}.

\begin{table}[!t]
\renewcommand{\arraystretch}{1.7}
\centering
\resizebox{\columnwidth}{!}{
\begin{tabular}{|c|cccccccccc|}
\hline 
Decay  & $a_9$ & $b_{9}$ & $a_{10}$ & $b_{10}$ & $a_S$ & $b_S$ & $a_P$ & $b_P$ & $c_{9S}$ & $c_{10P}$ \\ \hline\hline
$\Lambda_b\to n e^- \mu^+$ & $0.54(8)$ & $0.77(12)$ & $0.54(8)$ & $0.77(12)$ & $0.50(8)$ & $0.26(4)$ & $0.50(8)$ &  $0.26(4)$ & $-0.06(1)$ & $0.06(1)$  \\
$\Lambda_b\to n e^- \tau^+$ & $0.36(4)$  & $0.54(8)$ & $0.36(4)$  & $0.54(8)$& $0.29(4)$ & $0.13(2)$ & $0.29(4)$& $0.13(2)$  &$-0.47(3)$ & $0.47(3)$ \\ 
$\Lambda_b\to n \mu^- \tau^+$ & $0.35(4)$ & $0.55(8)$ & $0.34(4)$&  $0.55(8)$ & $0.28(4)$ & $0.13(2)$ & $0.30(4)$ & $0.14(2)$ 
&$-0.41(3)$ & $0.50(3)$\\ \hline

${{\mathit \Xi}^{0}}\to \Lambda e^- \mu^+$ & $0.33$ & $0.49$ & $0.33$ & $0.49$ & $0.36$ & $0.14$ & $0.36$ & $0.14$ & $-0.02$ & $0.02$  \\
${{\mathit \Xi}^{0}}\to \Lambda e^- \tau^+$ & $0.26$  & $0.32$ & $0.26$  & $0.32$& $0.22$ & $0.08$ & $0.22$ & $0.08$  &$-0.21$ & $0.21$ \\ 
${{\mathit \Xi}^{0}}\to \Lambda \mu^- \tau^+$ & $0.25$ & $0.34$ & $0.24$&  $0.34$ & $0.21$ & $0.08$ & $0.23$ & $0.08$ 
&$-0.19$ & $0.22$\\ \hline\hline

$\Lambda_b\to \Lambda e^- \mu^+$ &  $13.2(18)$ & $18.5(30)$ & $13.2(18)$& $18.5(30)$ & $12.3(18)$ &     $6.2(9)$ & $12.3(18)$ & $6.2(9)$ & $-1.4(1)$ & $1.4(1)$  \\
$\Lambda_b\to \Lambda e^- \tau^+$ & $8.6(9)$ & $13.3(17)$  & $8.6(9)$ & $13.3(17)$ & $7.0(10)$ & $3.2(4)$ & $7.0(10)$ & $3.2(4)$ & $-11.2(7)$ & $11.2(7)$ \\ 
$\Lambda_b\to \Lambda \mu^- \tau^+$ &  $8.5(9)$ & $13.6(2)$ & $8.4(9)$ & $13.8(2)$ & $6.7(9)$ &  $3.1(4)$ & $7.2(10)$  & $3.3(4)$ & $-9.7(6)$& $11.3(6)$  \\ \hline
\end{tabular}}
\caption{\small \sl Numerical coefficients entering Eq.~\eqref{eq:B-baryon} for the LFV baryon decays. For their evaluation in the $\Lambda_b\to n$ and $\Xi_b\to \Lambda$ cases, both based on $b\to d\ell_i^-\ell_j^+$, we respectively rely on the form factors computed in Refs.\cite{Detmold:2015aaa} and~\cite{Faustov:2018ahb}, while in the case of $\Lambda_b\to \Lambda$, based on $b\to s\ell_i^-\ell_j^+$, we used the form factors of Ref.~\cite{Detmold:2016pkz}. $\Xi_b\to\Lambda$ coefficients have no error bars because they are evaluated by using Ref.~\cite{Faustov:2018ahb}. Note also that the coefficients for the channels with the opposite lepton charges in the final state are the same as those listed above, except for $c_{9S}\to -c_{9S}$. All coefficients are to be multiplied by $10^{-9}$, cf.~Eq.~\eqref{eq:B-baryon}.}
\label{tab:magic-numbers-baryons} 
\end{table}

\subsection{Numerical significance}
\label{ssec:baryons-numerical}

We emphasize that the form factors used to describe 
$\Xi_b\to \Lambda \ell_i\ell_j$ are obtained in the framework of a quark model~\cite{Faustov:2018ahb} which is why why the corresponding entries in Table~\ref{tab:magic-numbers-baryons} have no uncertainties. Notice also that while in the case of the meson decays some of the magic numbers are strictly zero ($b_{S,P,9,10}$ in the $B\to P$ case, and $a_{S,P}$ in $B\to V$), in the case of baryon decays all of them are different from zero, which is another reason why the decays of baryons are valuable.

We can now use Eqs.~(\ref{eq:B-lep},\ref{eq:B-semilep},\ref{eq:B-baryon}) and the {\it magic numbers} given in Tables~\ref{tab:magic-numbers-mesons},\ref{tab:magic-numbers-mesons2},\ref{tab:magic-numbers-baryons} and compare the baryonic LFV modes to the leptonic and semileptonic meson decays discussed in Sec.~\ref{ssec:B-reminder}. As it has been previously discussed for the $b\to s$ transition in Ref.~\cite{Becirevic:2016zri,Becirevic:2016oho}, there is a hierarchy among the branching fractions when a scalar- or a vector- type of interaction is involved. Moreover, the hierarchy is distinct in these two different situations.  
\begin{itemize}
    \item[$\bullet$] If $C_{S,P}\neq 0$ and $C_{9,10}= 0$ there is a following hierarchy of the $b\to s$ decay modes:
\begin{align}
\label{eq:hierarchy-baryon-1}
\mathcal{B}(B_s\to \ell_i\ell_j)&>\mathcal{B}(B\to K \ell_i\ell_j)>\mathcal{B}(\Lambda_b \to \Lambda \ell_i\ell_j)>\mathcal{B}(B\to K^\ast \ell_i\ell_j)\,,
\end{align}

\noindent which remains the same for any flavor $i\neq j$. Similarly, for the $b\to d$ decays, we see that, 
\begin{align}
\label{eq:hierarchy-baryon-1-bis}
\mathcal{B}(B_d\to \ell_i\ell_j )&>\mathcal{B}(B\to \pi \ell_i\ell_j)>\mathcal{B}(\Lambda_b \to n \ell_i\ell_j)> \mathcal{B}(B\to \rho \ell_i\ell_j)\,.
\end{align}

\noindent where the leptonic decays have the largest branching fraction due to absence of helicity suppression. For instance, by assuming $C_S=-C_P\neq 0$, and after setting the primed coefficients to zero, we find that 
\begin{align}
\dfrac{\mathcal{B}(B\to K \ell_i\ell_j)}{\mathcal{B}(B_s\to \ell_i\ell_j)} &\simeq \left\{\begin{matrix}
0.13\,~\text{for }e\mu\\[0.3em] 
0.09\,,~\text{for }e\tau\\[0.3em]  
0.09\,,~\text{for }\mu\tau
\end{matrix}\right.\ ,&\quad
\dfrac{\mathcal{B}(B\to \pi\ell_i\ell_j)}{\mathcal{B}(B\to \ell_i\ell_j)} &\simeq \left\{\begin{matrix}
0.17\,,~\text{for }e\mu\\[0.3em] 
0.13\,,~\text{for }e\tau\\[0.3em]  
0.13\,,~\text{for }\mu\tau
\end{matrix}\right.\ ,\nonumber\\[0.75em]
\label{eq:ratio-scalar}
\dfrac{\mathcal{B}(B\to K^\ast \ell_i\ell_j)}{\mathcal{B}(B_s\to \ell_i\ell_j)} &\simeq \left\{\begin{matrix}
0.04\,,~\text{for }e\mu\\[0.3em] 
0.03\,,~\text{for }e\tau\\[0.3em]  
 0.03\,,~\text{for }\mu\tau
\end{matrix}\right.\ , &\quad
\dfrac{\mathcal{B}(B\to \rho\ell_i\ell_j)}{\mathcal{B}(B\to \ell_i\ell_j)} &\simeq \left\{\begin{matrix}
0.08\,,~\text{for }e\mu\\[0.3em] 
  0.04\,,~\text{for }e\tau\\[0.3em]  
0.05\,,~\text{for }\mu\tau
\end{matrix}\right. \ ,\\[0.75em]
\dfrac{\mathcal{B}(\Lambda_b\to \Lambda \ell_i\ell_j)}{\mathcal{B}(B_s\to \ell_i\ell_j)} &\simeq \left\{\begin{matrix}
 0.09\,,~\text{for }e\mu\\[0.3em] 
0.06\,,~\text{for }e\tau\\[0.3em]  
0.06\,,~\text{for }\mu\tau
\end{matrix}\right. \ ,&\quad
\dfrac{\mathcal{B}(\Lambda_b\to n\ell_i\ell_j)}{\mathcal{B}(B\to \ell_i\ell_j)} &\simeq \left\{\begin{matrix}
    0.12\,,~\text{for }e\mu\\[0.3em] 
0.08\,,~\text{for }e\tau\\[0.3em]  
 0.09\,,~\text{for }\mu\tau
\end{matrix}\right.\ , \nonumber
\end{align}
where we normalized by the largest branching fraction in this scenario. Note also that the remaining Wilson coefficients cancel out in these ratios. For simplicity, we only show the ratios' central values but the uncertainties due to form factors can be easily computed by using Tables~\ref{tab:magic-numbers-mesons},\ref{tab:magic-numbers-mesons2},\ref{tab:magic-numbers-baryons}.

\

    \item[$\bullet$] In contrast to the previous case, we now assume $C_{9,10}\neq 0$ and $C_{S,P} = 0$. The hierarchy pattern observed above becomes reversed, and for the $b\to s$ modes we find,
\begin{align}
\label{eq:hierarchy-baryon-2}
\mathcal{B}(B_s\to \ell_i\ell_j)<\mathcal{B}(B\to K \ell_1\ell_2)<\mathcal{B}(\Lambda_b \to \Lambda \ell_i\ell_j)<\mathcal{B}(B\to K^\ast \ell_i\ell_j)\,.
\end{align}

\noindent which also holds true for the $b\to d$ modes,
\begin{align}
\label{eq:hierarchy-baryon-2-bis}
\mathcal{B}(B\to \ell_i\ell_j)<\mathcal{B}(B\to \pi \ell_1\ell_2)<\mathcal{B}(\Lambda_b \to n \ell_i\ell_j)<\mathcal{B}(B\to \rho \ell_i\ell_j)\,.
\end{align}

\noindent To illustrate numerically this feature we now assume the currents governing LFV to be left-handed, which means $C_9=-C_{10}\neq 0$ while keeping $C_{9,10}^\prime = 0$, and we find:
\begin{align}
\dfrac{\mathcal{B}(B\to K \ell_i\ell_j)}{\mathcal{B}(B\to K^\ast \ell_i\ell_j)} &\simeq \left\{\begin{matrix}
 0.5\,,~\text{for }e\mu\\[0.3em] 
0.6\,,~\text{for }e\tau\\[0.3em]  
0.6\,,~\text{for }\mu\tau
\end{matrix}\right.\,,\nonumber&
\dfrac{\mathcal{B}(B\to \pi\ell_i\ell_j)}{\mathcal{B}(B\to \rho\ell_i\ell_j)} &\simeq \left\{\begin{matrix}
0.4\,,~\text{for }e\mu\\[0.3em] 
0.6\,,~\text{for }e\tau\\[0.3em]  
0.5\,,~\text{for }\mu\tau
\end{matrix}\right.\,,\\[0.75em]
\dfrac{\mathcal{B}(\Lambda_b\to \Lambda \ell_i\ell_j)}{\mathcal{B}(B\to K^\ast \ell_i\ell_j)} &\simeq \left\{\begin{matrix}
0.8\,,~\text{for }e\mu\\[0.3em] 
1.1\,,~\text{for }e\tau\\[0.3em]  
1.1\,,~\text{for }\mu\tau
\end{matrix}\right.\,,&
\dfrac{\mathcal{B}(\Lambda_b\to n \ell_i\ell_j)}{\mathcal{B}(B\to \rho\ell_i\ell_j)} &\simeq \left\{\begin{matrix}
0.9\,,~\text{for }e\mu\\[0.3em] 
1.1\,,~\text{for }e\tau\\[0.3em]  
 1.1\,,~\text{for }\mu\tau
\end{matrix}\right.\,,\\[0.75em]
\dfrac{\mathcal{B}(B_s\to \ell_i\ell_j)}{\mathcal{B}(B\to K^\ast \ell_i\ell_j)} &\simeq \left\{\begin{matrix}
1.4 \times10^{-3}\,,~\text{for }e\mu\\[0.3em] 
0.6\,,~\text{for }e\tau\\[0.3em]  
0.6\,,~\text{for }\mu\tau
\end{matrix}\right.\,,&
\dfrac{\mathcal{B}(B\to \ell_i\ell_j)}{\mathcal{B}(B\to \rho\ell_i\ell_j)} &\simeq \left\{\begin{matrix}
 1.1\times10^{-3}\,,~\text{for }e\mu\\[0.3em] 
 0.4\,,~\text{for }e\tau\\[0.3em]  
 0.4\,,~\text{for }\mu\tau
\end{matrix}\right.\nonumber\,.
\end{align}
\noindent Note again that all the ratios above are independent on the value of $C_9=-C_{10}\neq 0$. 

\end{itemize}

\begin{figure}[t!]
\centering
\includegraphics[width=0.5\linewidth]{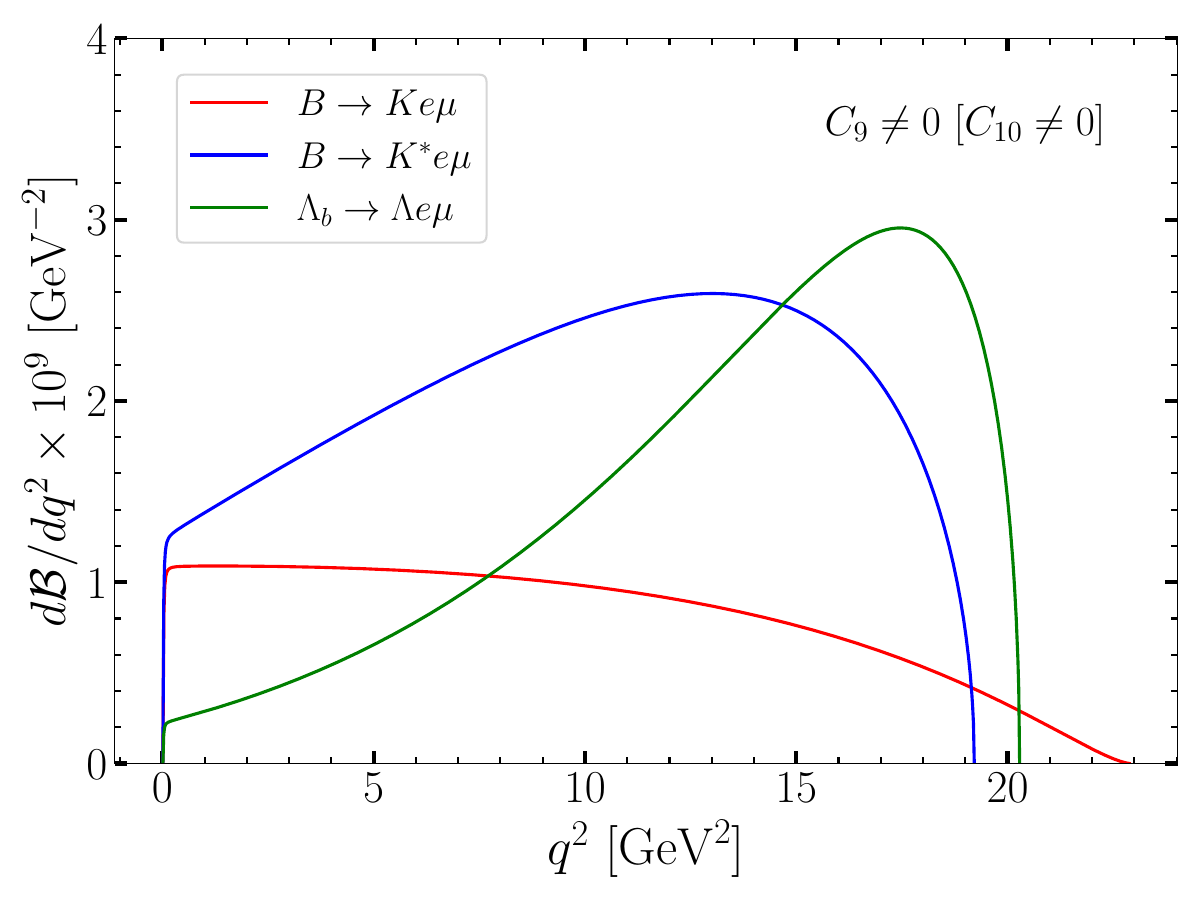}~\includegraphics[width=0.5\linewidth]{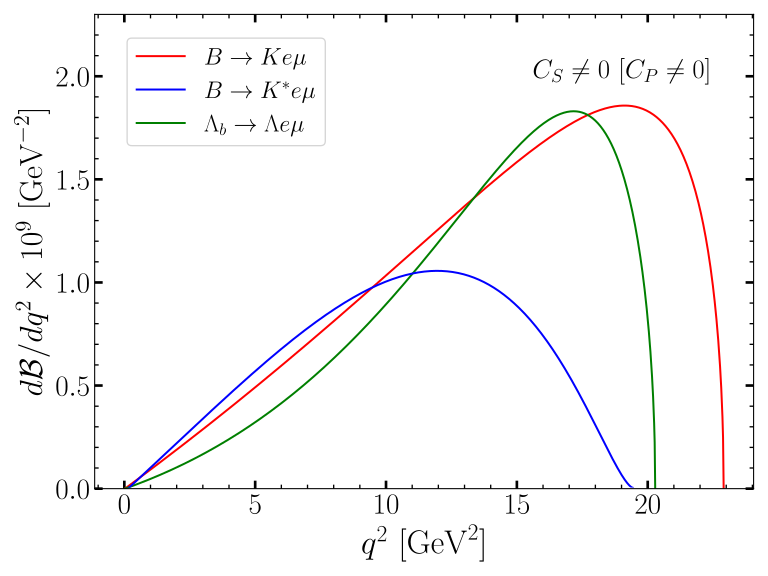}\\[0.5em]
\includegraphics[width=0.5\linewidth]{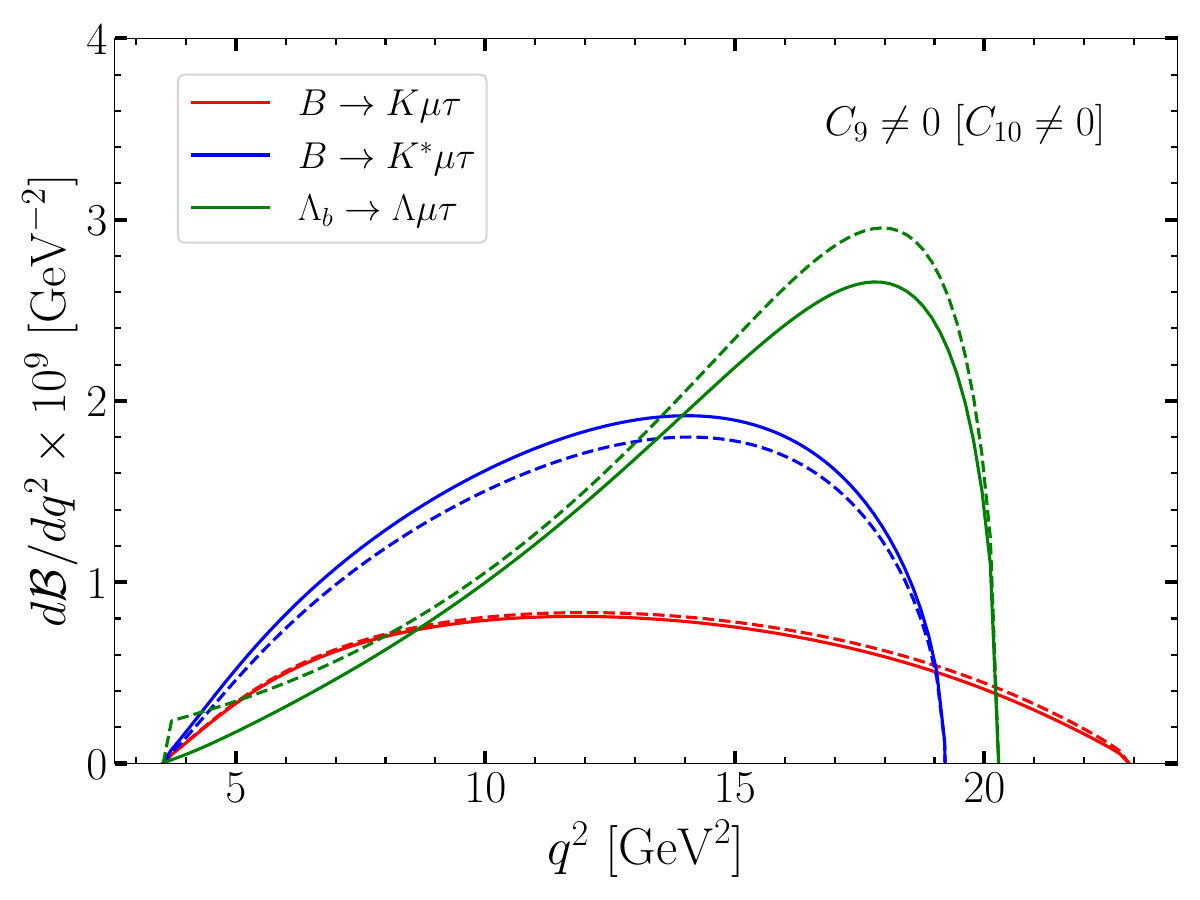}~\includegraphics[width=0.5\linewidth]{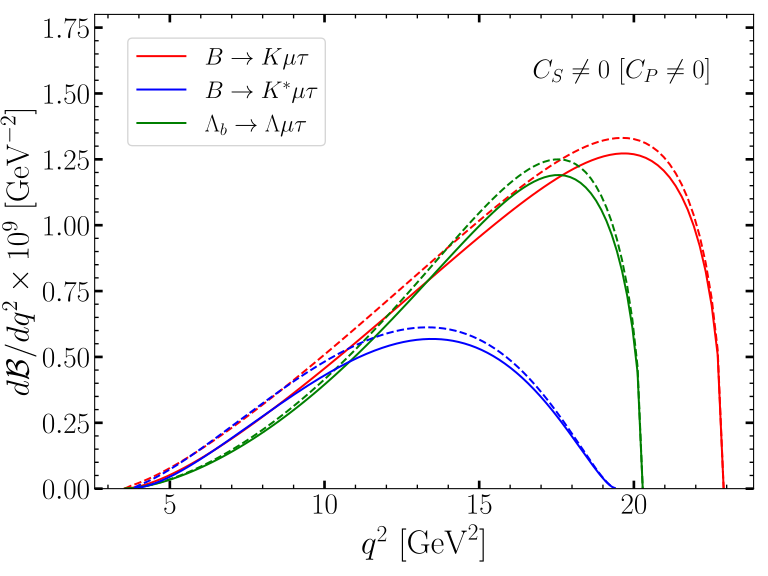}
\caption{\small \sl Differential distributions of $B\to K \ell_i \ell_j$ (red), $B\to K^\ast \ell_i \ell_j$ (blue) and $\Lambda_b \to \Lambda \ell_i \ell_j$ (green) decay rates (branching fractions), plotted as functions of the di-lepton invariant-mass ($q^2$) for the $b\to s \mu e$ (upper row) and $b\to s \mu \tau$ (lower row) transitions. Each curve is obtained by fixing a single effective coefficient to $\mathcal{C}=1$, with the others set to zero. The distributions for $C_{9}$ (solid line) and $C_{10}$ (dashed line) are shown in the left plots, whereas the ones for $C_{S}$ (solid line) and $C_{P}$ (dashed line) are displayed in the right ones. Distributions of decays to $e\tau$ are not shown because they are very similar to the $\mu\tau$ ones.}
\label{fig:diff-q2-bs} 
\end{figure}

\subsection{Differential distributions}
\label{ssec:diff}

The $q^2$-distribution of LFV decays can be useful when trying to distinguish among various EFT scenarios~\cite{Becirevic:2016zri}. Such information is an important input for the Monte Carlo simulations of the signals in experimental analyses, which can yield different experimental bounds depending on the type of EFT scenario considered. This type of information was used in the recent LHCb study~\cite{LHCb:2020khb} as well as in the Belle~\cite{Belle:2022pcr} searches of $B\to K\mu\tau$. 

\begin{figure}[t!]
\centering
\includegraphics[width=0.5\linewidth]{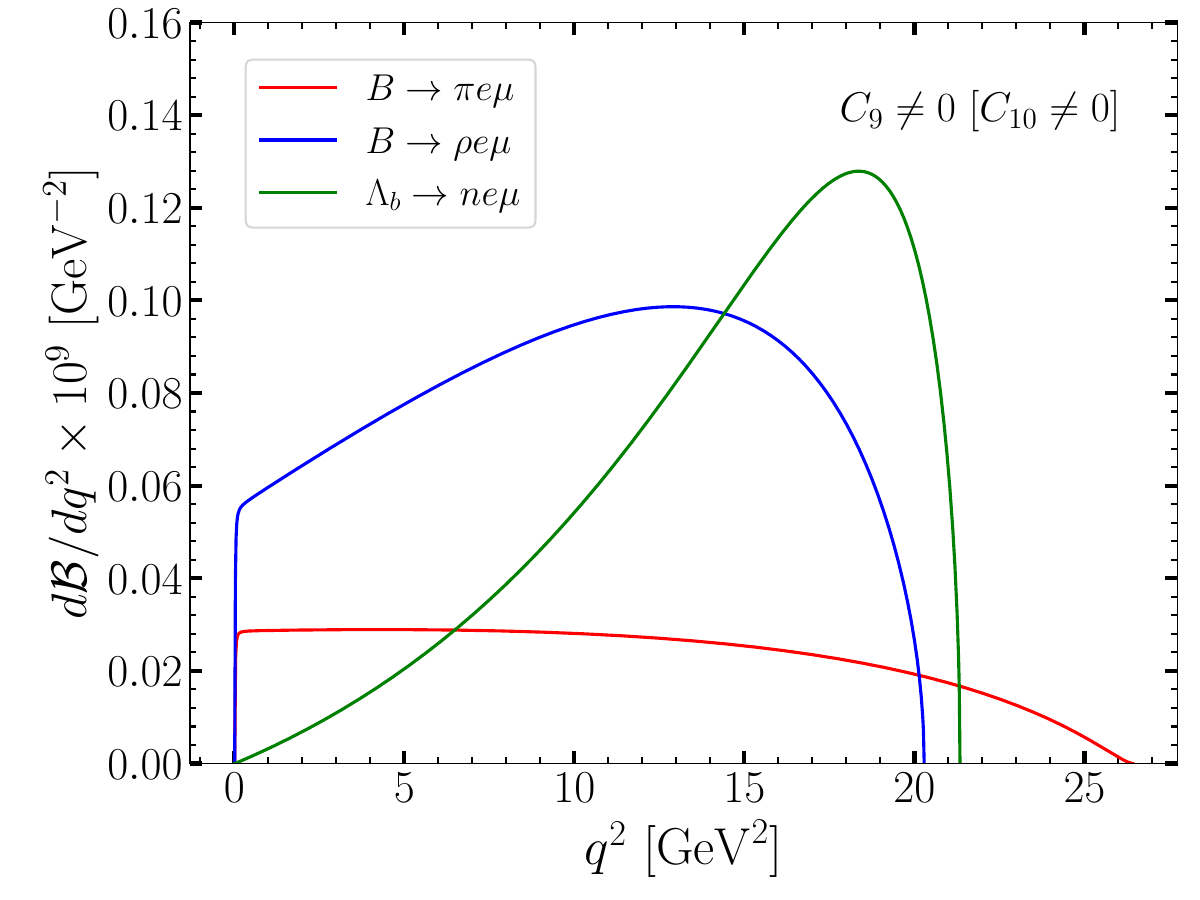}~\includegraphics[width=0.5\linewidth]{Figs/cScPSkin_emu_d.pdf}\\[0.5em]
\includegraphics[width=0.5\linewidth]{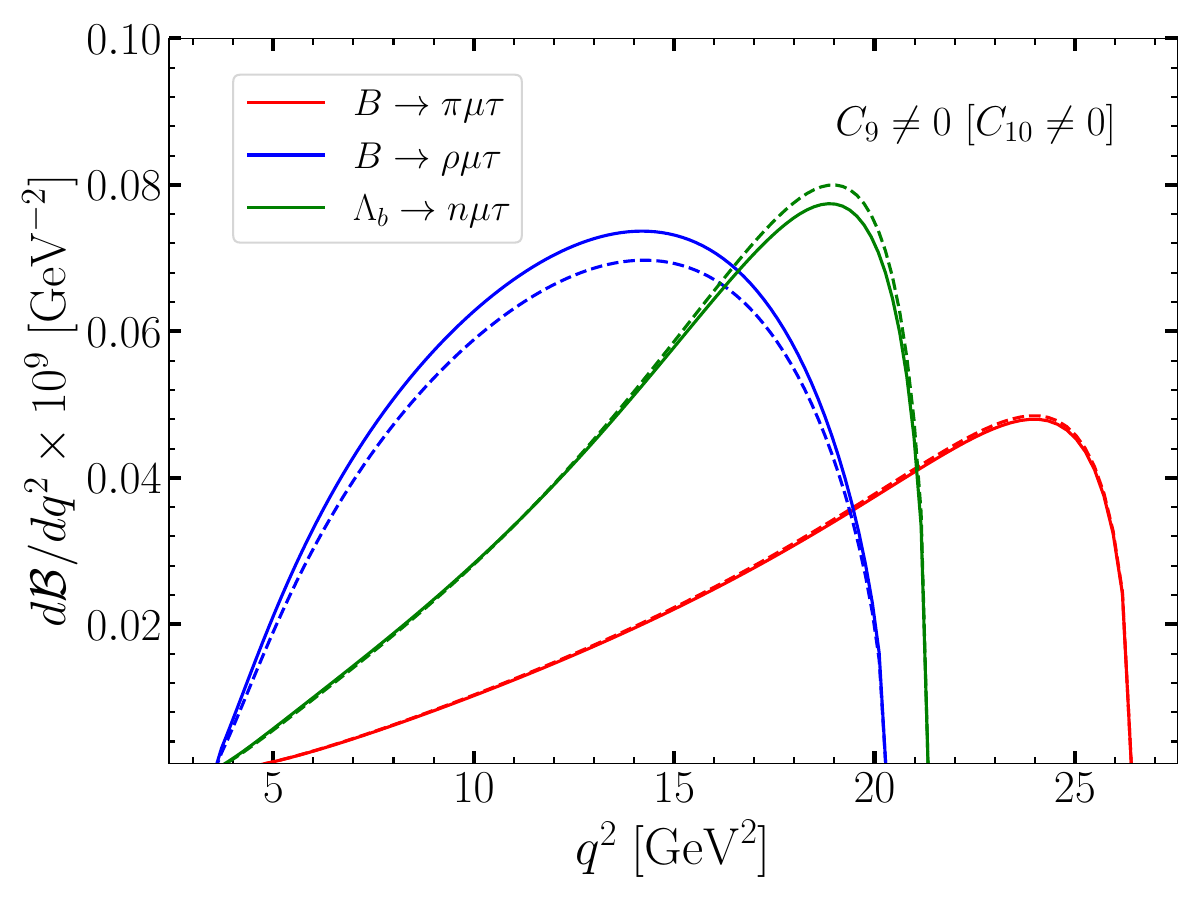}~\includegraphics[width=0.5\linewidth]{Figs/cscpskin_2.pdf}
\caption{\small \sl Differential distributions of $B\to \pi \ell_i \ell_j$ (red), $B\to \rho \ell_i \ell_j$ (blue) and $\Lambda_b \to n \ell_i \ell_j$ (green) are plotted as  functions of the di-lepton invariant-mass ($q^2$) for the $b\to d \mu e$ (upper row) and $b\to d \mu \tau$ (lower row). N.B. that the decays to $e\tau$ are not shown as they are indistinguishable from the ones to $\mu\tau$. }
\label{fig:diff-q2-bd} 
\end{figure}

In Fig.~\ref{fig:diff-q2-bs}, we show the differential distributions of $B\to K^{(\ast)}\ell_i\ell_j$ and $\Lambda_b\to \Lambda\ell_i\ell_j$ decays for vector (left column) and scalar operators (right column) into both $\mu e$ (upper row) and $\tau \mu$ channels (lower row).~\footnote{Predictions for the decays into $\tau e$ are not shown as they are very similar to the $\tau \mu$ ones.} The curves related to the meson decays should be viewed as an update of those provided in Ref.~\cite{Becirevic:2016zri}. For the vector coefficients, we find that the $q^2$-shapes of $B\to K$, $B\to K^\ast$, and $\Lambda_b \to \Lambda$ differ. Instead, for the scalar coefficients they all peak at high $q^2$-values, as it can be seen in the right plots of Fig.~\ref{fig:diff-q2-bs}. Therefore, if LFV is indeed observed in these channels, this difference could be used to disentangle among the operators contributing to these decays. In Fig.~\ref{fig:diff-q2-bd}, we show the analogous plots for the $b\to d$ decays, namely $B\to \pi \ell_i\ell_j$, $B\to \rho\ell_i\ell_j$ and $\Lambda_b\to n\ell_i\ell_j$. The main difference with the previous case is the larger phase space that makes the difference between scalar and vector operators even more evident, especially for $B\to \pi$.

\subsection{Indirect bounds on $\Lambda_b$ decays}
\label{ssec:indirect}

Even though the LFV decays of a $b$-flavored baryon have not yet been searched experimentally, we can use the available information on the similar leptonic and semileptonic meson decays to constrain their LFV branching ratios. To this purpose we rely on the EFT Lagrangian~\eqref{eq:hamiltonian} and write the LFV baryon branching fractions as a quadratic form made with vectors made of Wilson coefficients, to wit,
\begin{align}
    \mathcal{B}(\Lambda_b\to\Lambda\ell_i\ell_j) \equiv \vec{v}^{\,\dagger}\cdot  M_{\Lambda}^{ij} \cdot \vec{v}\,,
\end{align}
with $\vec{v}$ being a vector of Wilson coefficients at $\mu=m_b$,
\begin{align}
    \vec{v}= \Big{(}
    C_{9}^{(+)}\,,C_{9}^{(-)}\,,
    C_{10}^{(+)}\,,C_{10}^{(-)}\,,
    C_{S}^{(+)}\,,C_{S}^{(-)}\,,
    C_{P}^{(+)}\,,C_{P}^{(-)}
    \Big{)}^{\,T}\,,
\end{align}
where $C_I^{(\pm)} \equiv  C_I \pm C_{I^\prime}$, with $I\in \lbrace 9,10,S,P\rbrace$, and the flavor indices are omitted for simplicity. The matrix $M_\Lambda \equiv M_{\Lambda_b\to\Lambda\ell_i\ell_j}$ is positive definite, with the entries given in Eq.~\eqref{eq:Lb-semilep}. Similar quadratic forms can be written for $\mathcal{B}(B_s\to\ell_i\ell_j)$ and $\mathcal{B}(B\to K^{(\ast)}\ell_i\ell_j)$,  associated to $M_{B_s}^{ij}$ and $\smash{M_{K^{(\ast)}}^{ij}}$ respectively, but these matrices are only positive semi-definite, due to the zero entries that can be seen, e.g.,~in Table~\ref{tab:magic-numbers-mesons}. By varying the effective coefficients, we can then determine the largest possible values of baryonic rates that are compatible with the existing constraints on meson decays, cf. Table~\ref{tab:exp},~\footnote{Note that this problem can be re-expressed as an eigenvalue problem and then solved semi-analytically, as discussed in Ref.~\cite{Descotes-Genon:2023pen} in a slightly different context.} 
\begin{align}
\label{eq:max-br-eft}
    \mathcal{B}(\Lambda_b\to\Lambda\ell_i\ell_j)_\mathrm{max}= \max_{v} \Big{\lbrace} \vec{v}^{\,\dagger}\cdot  M_{\Lambda}^{ij} \cdot \vec{v} \;\;\big{|}\;\;\mathrm{for}~\;~&\vec{v}^{\,\dagger}\cdot  M_{B_s}^{ij}\cdot v\leq \mathcal{B}(B_s\to \ell_i\ell_j)^\mathrm{exp}\,,\\*[0.3em] \;\;\mathrm{and}~\;~&\vec{v}^{\,\dagger}\cdot  M_{K^{(\ast)}}^{ij} \cdot v\leq \mathcal{B}(B\to K^{(\ast)}\ell_i\ell_j)^\mathrm{exp}
    \Big{\rbrace} \,, \nonumber
\end{align}
This is a well-defined problem since leptonic and semileptonic decays can potentially cover all possible directions in parameter space, provided the experimental limits are available for three channels, namely~$B_s\to\ell_i\ell_j$, $B\to K\ell_i\ell_j$ and $B\to K^\ast\ell_i\ell_j$. We emphasize that the bounds obtained in this way are fully model independent, as we do not introduce any specific assumption regarding the Wilson coefficients. Our only requirement is that the scale of New Physics must be large so that these decays can be consistently described by EFT~\eqref{eq:hamiltonian}.

The indirect upper limits obtained using the method described above are listed in Table~\ref{tab:indirect-bounds-EFT}. We find the $\mathcal{O}(10^{-8})$ bounds for the $b\to s e\mu$ channel, and $\mathcal{O}(10^{-5})$ for $b\to s\mu\tau$. These ranges of branching ratios can perhaps be accessible by the LHCb experiment, as estimated, e.g.,~for the $\mu\tau$ channel in Ref.~\cite{Bordone:2021usz}. Note, in particular, that we could not derive a similar \emph{model independent} bound for the $\Lambda_b \to \Lambda e\tau$ decay, neither for the decay modes based on the $b\to d\ell_i\ell_j$ transition, since the experimental constraints on the relevant $P\to V\ell_i\ell_j$ modes are not available for these transitions. To derive upper bounds for these decays, we need to invoke further assumptions on the effective coefficients, as we explore in the next Section, within concrete scenarios.

\begin{table}[!t]
\small
\renewcommand{\arraystretch}{1.6}
\centering
\begin{tabular}{|c|c||c|c|}
\hline 
Decay  & Indirect EFT bound & Decay  & Indirect EFT bound  \\
\hline\hline
$\Lambda_b\to \Lambda e^- \mu^+$ & $\lesssim 7\times10^{-9} $& $\Lambda_b\to n  e^- \mu^+$ & -- \\ 
$\Lambda_b\to \Lambda e^- \tau^+$ &  -- & $\Lambda_b\to n  e^- \tau^+$ & -- \\ 
$\Lambda_b\to \Lambda \mu^- \tau^+$ & $\lesssim 1.9\times10^{-5} $ & $\Lambda_b\to n  \mu^- \tau^+$ & -- \\ \hline
\end{tabular}
\caption{\small \sl Indirect EFT bounds obtained for each decay mode by using the available $95\%$~CL experimental bounds on the (semi)leptonic decays collected in Table~\ref{tab:exp}, as described in Eq.~\eqref{eq:max-br-eft}. The dashes indicate that a model-independent limit cannot be obtained for a specific channel since there are not enough experimental constraints at low energies.}
\label{tab:indirect-bounds-EFT} 
\end{table}

\section{Concrete scenarios}
\label{sec:concrete}

So far we dealt with the EFT approach without specifying the New Physics scenario. We now turn to two concrete models that can generate the LFV operators: (i) a model with LFV couplings to the Higgs boson, and (ii) a scenario with a heavy $Z^\prime$.~\footnote{ Concrete scenarios with $\mathcal{O}(1~\mathrm{TeV})$ scalar leptoquarks can also induce large LFV rates for $b\to q\ell_i\ell_j$, as studied e.g.~in Ref.~\cite{Becirevic:2016oho,Bordone:2018nbg,Cornella:2021sby,Crivellin:2019dwb}.} We will apply additional constraints that imply stronger limits than those shown in Table~\ref{tab:indirect-bounds-EFT}, derived by relying on EFT alone. Furthermore, we will discuss the impact of the LHC searches for LFV such as $h\to\ell_i\ell_j$ and $pp\to\ell_i\ell_j$ at high-$p_T$.

\subsection{Higgs LFV-couplings: $C_{S(P)}\neq 0$ }
\label{ssec:higgs}

New Physics contributions to the Higgs boson coupling to leptons can be described by the following dimension-six operator~\cite{Harnik:2012pb}:
\begin{equation}
    \mathcal{L}^{(6)}_\mathrm{eff}  \supset \dfrac{c_{eH}^{ij}}{\Lambda^2}\,\big{(}H^\dagger H\big{)} \big{(}\overline{L}_i H e_{Rj}\big{)}+\mathrm{h.c.}\,,
\end{equation}

\noindent where $L$ and $e_R$ denote the $SU(2)_L$ doublet and singlet fields, respectively, $H$ is the SM Higgs doublet, and $c_{eH}^{ij}$ are the effective coefficients associated to the lepton flavors $i,j$. After spontaneous electroweak symmetry breaking, the Yukawa coupling to charged leptons is given by $\mathcal{L}_{\mathrm{Yuk}} \supset  -y_{\ell}^{ij}\,h\, \bar{\ell}_{Li}\ell_{Rj}+\mathrm{h.c.}$, where
\begin{equation}
    y_\ell^{ij}= \delta_{ij}\, \dfrac{m_{\ell_i}}{v}- \dfrac{v^2}{\sqrt{2} \Lambda^2} c_{eH}^{ij}\,,
\end{equation}

\begin{figure}[t!]
\centering
\includegraphics[width=0.485\linewidth]{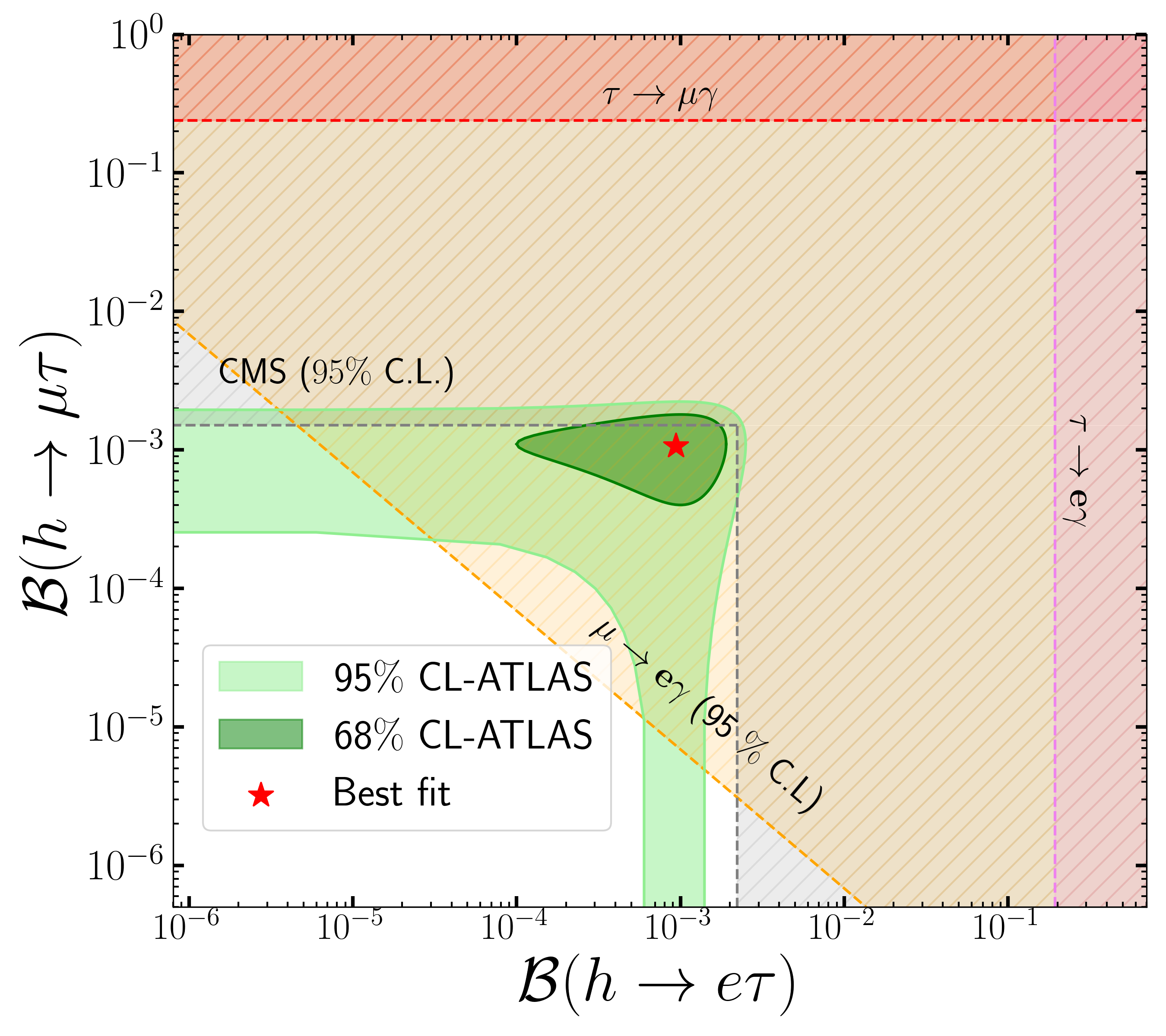}~\quad~\includegraphics[width=0.485\linewidth]{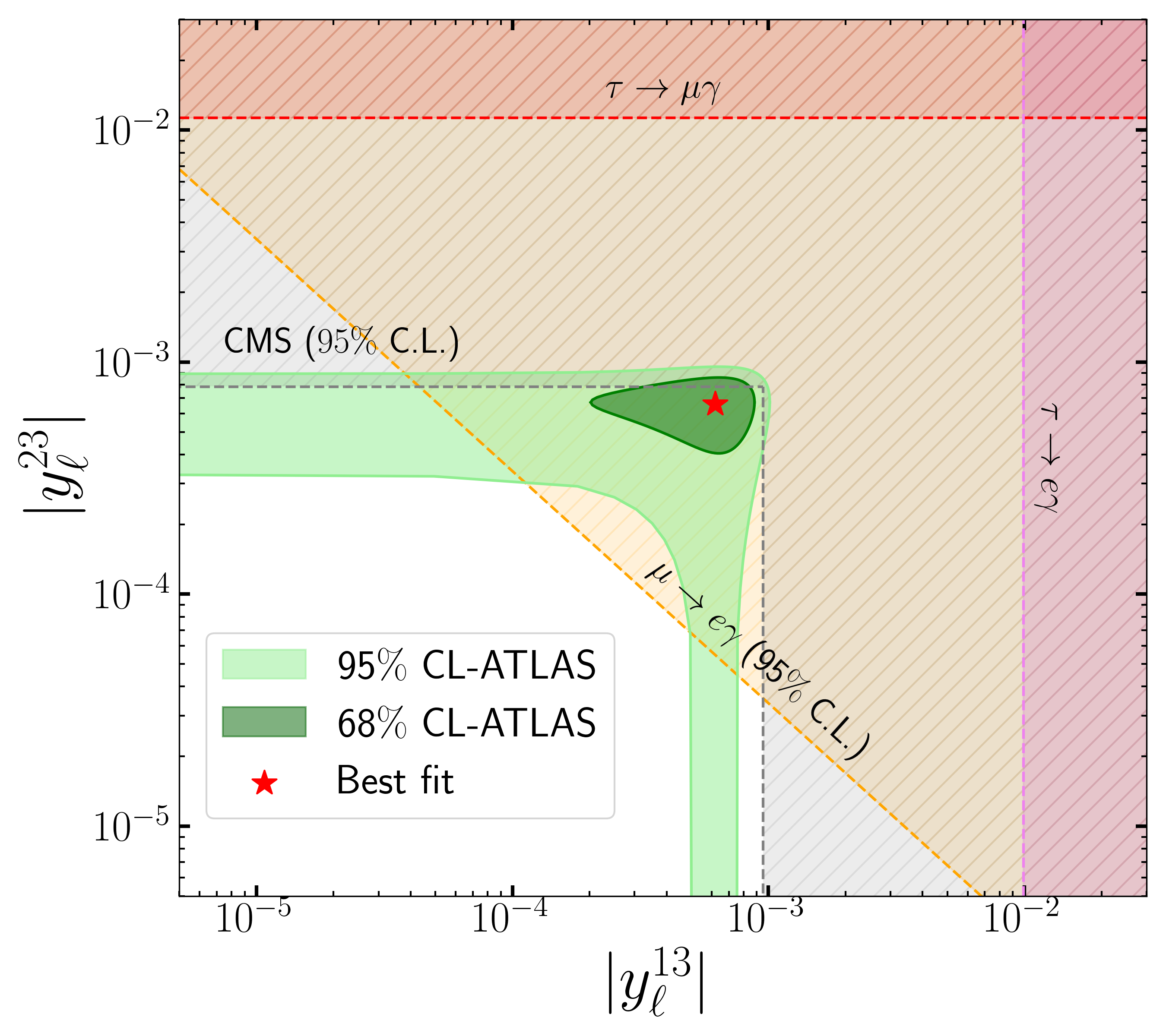}
\caption{\small \sl In the left panel, we display $\mathcal{B}(h\to \mu \tau)$ vs $\mathcal{B}(h\to e \tau)$. The values preferred by ATLAS (green) \cite{ATLAS:2023mvd} are compared to the CMS bounds \cite{CMS:2021rsq} (gray), and to the indirect bounds deduced from $\mu\to e\gamma$ (orange), $\tau\to e\gamma$ (magenta) and $\tau\to\mu\gamma$ (red) decays. In the right panel, we show the allowed regions of $|y^{23}_\ell|$ vs.~$|y^{23}_\ell|$ plane, by assuming $y_\ell$ to be a Hermitian matrix. Note that the indirect bounds actually exclude the best fit point value reported by ATLAS (red star).}
\label{fig:higgs-lfv-plot} 
\end{figure}

\noindent with $v=(\sqrt{2}G_F)^{-1/2}$ being the Higgs vacuum expectation value, and we assume that $c_{eH}$ is real. This coupling contributes at one-loop to the $b\to q \ell_i\ell_j$ transition via Higgs-penguin diagrams, which induce the (pseudo)scalar effective coefficients~\cite{Li:2014fea}:
\begin{align}
\begin{split}
    C_{S(P)}^{(q)\,ij} & = - \dfrac{y_{\ell}^{ij}\pm y_{\ell}^{ji\,\ast}}{2} \dfrac{m_b}{v} \dfrac{1}{16\pi \alpha_\mathrm{em}} \\
    &\times\bigg{[} \dfrac{6x_t}{x_h} - \dfrac{2x_t^3}{(1-x_t)^3}\log x_t + \dfrac{4 x_t^2}{(1-x_t)^3 }\log x_t - \dfrac{x_t^2}{(1-x_t)^2} + \dfrac{3x_t}{(1-x_t)^2}\bigg{]}\,,
\end{split}
\end{align}

\noindent where $x_{t,h}=m_{t,h}^2/m_W^2$. Note that only $C_S$ is generated if $y_\ell^{ij} = y_\ell^{ji\ast}$, i.e.,~if the matrix with $c_{eH}^{ij}$ entries is Hermitian.

LFV Higgs couplings $y_\ell^{ij}$ directly induce the $h\to \ell_i\ell_j$ decays that are currently studied experimentally~\cite{ATLAS:2023mvd,CMS:2021rsq}. The branching fractions for these decays are given by
\begin{align}
    \mathcal{B}(h\to\ell_i\ell_j) = \dfrac{m_h}{8\pi\Gamma_h} \big{(} |y_\ell^{ij}|^2+|y_\ell^{ji}|^2 \big{)}\,,
\end{align}

\noindent where $\Gamma_h \simeq 4.1$~MeV is the SM prediction of the Higgs total width~\cite{LHCHiggsCrossSectionWorkingGroup:2016ypw}. In the latest experimental search of $h\to e\tau$ and $h\to \mu\tau$ decays by ATLAS, a mild excess was found in the region where both decays are generated~\cite{ATLAS:2023mvd}. CMS, instead, reported upper limits on both decay modes~\cite{CMS:2021rsq}, as shown in Fig.~\ref{fig:higgs-lfv-plot}. 

\begin{figure}[t!]
\centering
\includegraphics[width=0.92\linewidth]{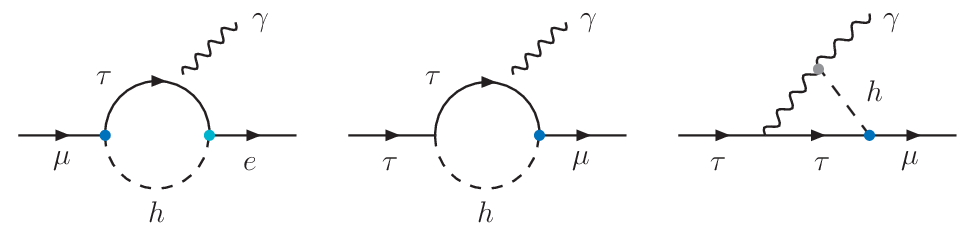}
\caption{\small \sl Loop contributions from the Higgs LFV couplings to the $\mu\to e \gamma$ (left diagram) and $\tau\to \mu \gamma$ decays (center and right diagrams) via a single and double insertion of the LFV couplings, respectively. The blue dots denote the EFT insertions, while the gray dot denotes the SM one-loop contribution to the $h\gamma\gamma$ vertex. }
\label{fig:higgs-lfv} 
\end{figure}

Importantly, however, the LFV Higgs couplings are also subject to indirect constraints from $\ell_i\to \ell_j \gamma$ decays: (i) via the Bar-Zee contributions to $\tau\to \ell \gamma$ (with $\ell=e,\mu$) that allow us to separately probe the $e\tau$ and $\mu\tau$ couplings, and (ii) via the insertions of both Yukawa couplings, generating the $\mu\to e\gamma$ process, as illustrated in Fig.~\ref{fig:higgs-lfv}. These effects are described by the effective dipole operators, 
\begin{align}
\label{eq:eff-dip}
\mathcal{L}_\mathrm{eff} \supset -\frac{C_{D_L}^{ij}}{m_{\ell_j}} \,\bar{\ell}_{Ri} \sigma_{\mu\nu} \ell_{Lj} \,F^{\mu\nu}-\frac{C_{D_R}^{ij}}{m_{\ell_j}} \bar{\ell}_{Li} \sigma_{\mu\nu} \ell_{Rj} \,F^{\mu\nu}
+ \mathrm{h.c.}\,,
\end{align}

\noindent $C_{D_{R(L)}}^{ij}$ are effective coefficients which depend on Yukawa couplings~\cite{Harnik:2012pb}. The corresponding branching fractions, assuming $i < j$, are given by:
\begin{align}
\mathcal{B}(\ell_j\to \ell_i\gamma)=  \tau_{\ell_j}\dfrac{m_{\ell_j} }{4\pi } \left(|C_{D_L}^{ij}|^2+|C_{D_R}^{ij}|^2\right)\,,
\end{align}
where $\tau_{\ell_j}$ is the lifetime of the decaying lepton. By using the current experimental limits, $\mathcal{B}(\tau\to e\gamma )<3.3\times 10^{-8}$ and $\mathcal{B}(\tau\to \mu\gamma )<4.2\times 10^{-8}$ at 90\%~CL~\cite{ParticleDataGroup:2022pth}, and the two-loop expressions from Ref.~\cite{Harnik:2012pb}, we obtain the indirect constraints on $\mathcal{B}(h\to \ell_i\ell_j)$, as shown in Fig.~\ref{fig:higgs-lfv-plot}. These constraints are induced by the last two diagrams in Fig.~\ref{fig:higgs-lfv} and they turn out to be considerably weaker than the direct limits from $h\to \ell_i \ell_j$. More importantly, however, the $\mu\to e\gamma$ decay is sensitive to the product of $y_{\ell}^{13}$ and $y_{\ell}^{23}$, described by the first diagram in Fig.~\ref{fig:higgs-lfv}. The corresponding contribution to  $C_{D_{L(R)}}^{e\mu}$ can be obtained by using the results of Refs.~\cite{Cornella:2019uxs} and it reads,
\begin{align}
    C_{D_{L(R)}}^{e\mu} = \dfrac{e\, y_\ell^{13}y_\ell^{23}}{64\pi^2}\dfrac{m_\mu m_\tau}{m_h^2}\left[3+2 \log (m_\tau^2/m_h^2)) \right]+\dots\,,
\end{align}
 
\noindent where we assumed $y_\ell^{ij} = y_\ell^{ji\ast}$, and neglected contributions proportional to $y_\ell^{12}$ as they are further suppressed.~\footnote{This is motivated by the chirality suppression of these contributions and by the stringent experimental limits on $\mathcal{B}(h\to e\mu)^\mathrm{exp}<4.4\times 10^{-5}$ (95\% CL)~\cite{CMS:2023pte}.} By combining these results, we find the following relation,
\begin{align}
\mathcal{B}(\mu\to e \gamma) \gtrsim 
6 \times 10^{-11}
\Bigg[ \frac{ \mathcal{B}(h\to e\tau) }{10^{-3}}\Bigg]
\Bigg[\frac{  \mathcal{B}(h\to \mu\tau)}{10^{-3}}\Bigg] \,,
\end{align}
which is then in manifest conflict with a strong experimental limit, $\mathcal{B}(\mu\to e\gamma)<3.1\times 10^{-13}$~\cite{MEG:2016leq}, when compared with the values preferred by ATLAS for the LFV Higgs decays to $1\sigma$, as it can be appreciated in Fig.~\ref{fig:higgs-lfv-plot}. 
In other words, it is not possible to simultaneously accommodate in this scenario the experimental $1\sigma$ region of $\mathcal{B}(h\to e\tau)$ and $\mathcal{B}(h\to\mu\tau)$ reported by ATLAS~\cite{ATLAS:2023mvd}, and the experimental bound on $\mathcal{B}(\mu\to e\gamma)$.~\footnote{See Ref.~\cite{Ardu:2022pzk} for a discussion of similar effects for other types of SMEFT operators.} This example also illustrates the complementarity of these indirect searches of LFV Higgs couplings and the direct ones since the low- and high-energy probes are sensitive to the same couplings.

Let us now focus on what the above discussion implies to the $b$-flavored baryon decays. To that end we can assume that only the $h\to e\tau$ coupling is present (i.e., set the $\mu\tau$ coupling to zero), and derive from Fig.~\ref{fig:higgs-lfv} that
\begin{align}
\sqrt{|y_\ell^{13}|^2+|y_\ell^{31}|^2}&< 1.4 \times 10^{-3}\,,
\end{align}
which then results in
\begin{align}
\mathcal{B}(\Lambda_b\to \Lambda e \tau)&< 4\times 10^{-15}\,.  
\end{align}

\noindent Conversely, if we set the $e\tau$ coupling to zero, from the current bound we get,
\begin{align}
\sqrt{|y_\ell^{23}|^2+|y_\ell^{32}|^2}&< 1.2\times 10^{-3}\,, 
\end{align}
which then translates to
\begin{align}
\mathcal{B}(\Lambda_b\to \Lambda \mu \tau)< 3\times10^{-15}\,,
\end{align}

\noindent showing that any LFV mode will have very small $\mathcal{B}(\Lambda_b\to \Lambda \ell_i\ell_j)$ in this type of scenarios, well beyond the reach of current experiments. The same conclusions hold true for the $B$-meson decays, which are related to the baryonic mode via Eq.~\eqref{eq:ratio-scalar}.

\subsection{LFV mediated by $Z^\prime$: $C_{9(10)}\neq 0$ }
\label{ssec:Zp}

Another simple scenario that could generate LFV in $\Lambda_b$ decays is the SM extended by a $Z^\prime \sim (\mathbf{1},\mathbf{1},0)$~\cite{Becirevic:2016zri,Crivellin:2015era}. In contrast to the previous model, in this case the LFV decay rates can be much larger since $Z^\prime$ contributes to $b\to d\ell_i\ell_j$ and $b\to s\ell_i\ell_j$ at tree level and the corresponding contribution is not suppressed by the $b$-quark mass. To specify the model we assume that $Z^\prime$ couples to the left-handed SM fermions as follows,
\begin{align}
\mathcal{L}_{Z^\prime} = 
-\dfrac{1}{4} Z^\prime_{\mu\nu} Z^{\prime\,\mu\nu} + \dfrac{m_{Z^\prime}^2}{2}Z_\mu^{\prime} Z^{\prime\,\mu}  +J_\mu Z^{\prime \mu}\,,
\quad \mathrm{with}\ \, J^\mu  = \sum_{\psi,i,j}g^{ij}_{L/R} \,\bar{\psi}_i \gamma^\mu P_{L/R} \psi_j\,,
\end{align}
where $Z^\prime_{\mu\nu}=D_\mu Z^\prime_\nu - D_\nu Z^\prime_\mu$, and $g$ denotes a coupling to the SM fermions, $\psi\in \lbrace Q, L, u_R,d_R,e_R\rbrace$. We will assume that the CKM matrix appears in the upper component of the quark doublet, $Q_i=[(V^\dagger\,u_{L})_i~d_{Li}]^T$. From  hermiticity of the above Lagrangian we know that $g^{ij}_{L/R} = g^{ji\,\ast}_{L/R}$. The above couplings give rise to the following Wilson coefficients relevant to $b\to s \ell_i\ell_j$:
\begin{align}
\label{eq:Zp-C10}
C_{9}^{s\,ij} &= -\dfrac{v^2}{m_{Z^\prime}^2}\dfrac{\pi}{\alpha_\mathrm{em} V_{tb}V_{ts}^\ast} g^{sb}_{L} (g^{ij}_{L}+g^{ij}_{R})\,, & \qquad C_{9^\prime}^{s\,ij} &= -\dfrac{v^2}{m_{Z^\prime}^2}\dfrac{\pi}{\alpha_\mathrm{em} V_{tb}V_{ts}^\ast} g^{sb}_{R} (g^{ij}_{L}+g^{ij}_{R})\,,  \\[0.4em]
\label{eq:Zp-C9}
C_{10}^{s\,ij} &= +\dfrac{v^2}{m_{Z^\prime}^2}\dfrac{\pi}{\alpha_\mathrm{em} V_{tb}V_{ts}^\ast} g^{sb}_{L} (g^{ij}_{L}-g^{ij}_{R})\,, & C_{10^\prime}^{s\,ij} &= +\dfrac{v^2}{m_{Z^\prime}^2}\dfrac{\pi}{\alpha_\mathrm{em} V_{tb}V_{ts}^\ast} g^{sb}_{R} (g^{ij}_{L}-g^{ij}_{R})\,, 
\end{align}
with analogous expressions for $b\to d \ell_i\ell_j$. In our analysis we consider two benchmark scenarios:~\footnote{The reader should not be confused by the notation adopted in this subsection, where for the quark superindices we use their flavors ``$sb$" or ``$db$", while for the leptons we use the superindices $ij$, with $i,j\in \{1,2,3\}$, corresponding to $\{e,\mu,\tau\}$, respectively.}
\begin{itemize}
    \item[$\bullet$]\textbf{Scenario I:} Only non-zero are the left-handed couplings to $Z^\prime$, i.e.~$C_9 = -C_{10}$,
    \item[$\bullet$]\textbf{Scenario II:} Only non-zero are the right-handed couplings to $Z^\prime$, i.e.~$C_{9^\prime} = +C_{10^\prime}$.
\end{itemize}
In order to constrain the $Z^\prime$ couplings one can use  several low energy observables as we describe in the following~\cite{Becirevic:2016zri,Crivellin:2015era}: 

\begin{figure}[t!]
\centering
\includegraphics[width=0.5\linewidth]{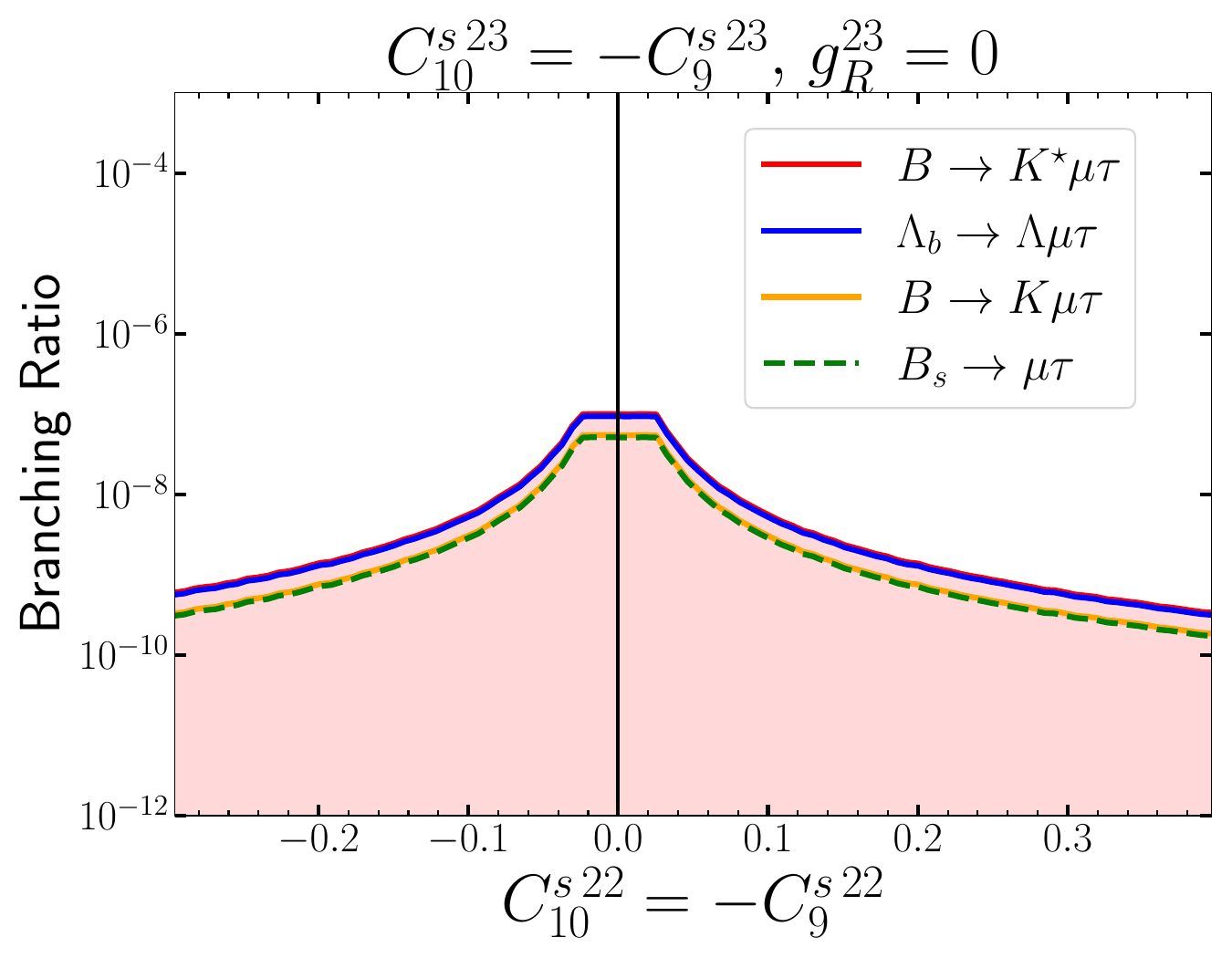}~\includegraphics[width=0.5\linewidth]{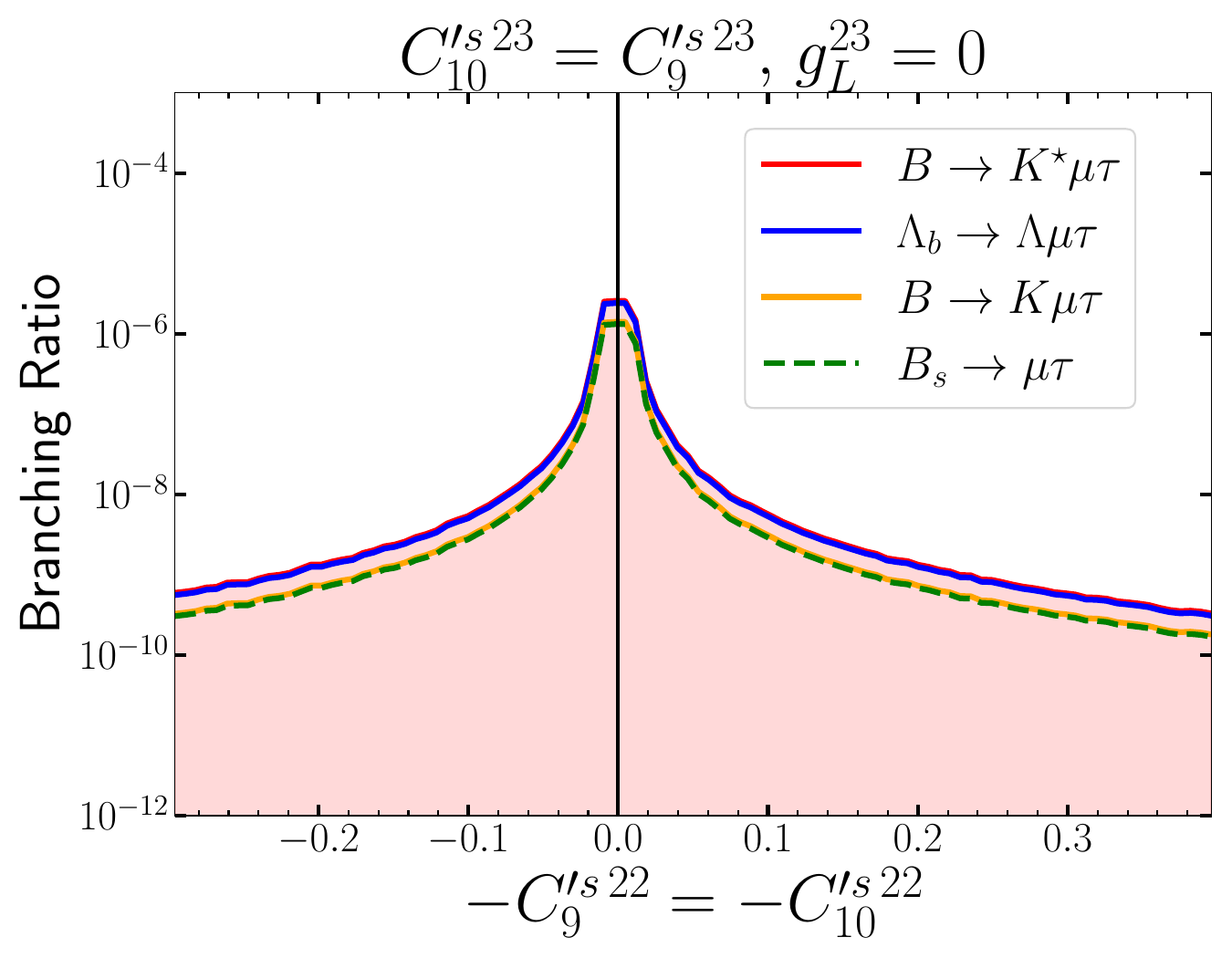}\\[0.5em]
\includegraphics[width=0.5\linewidth]{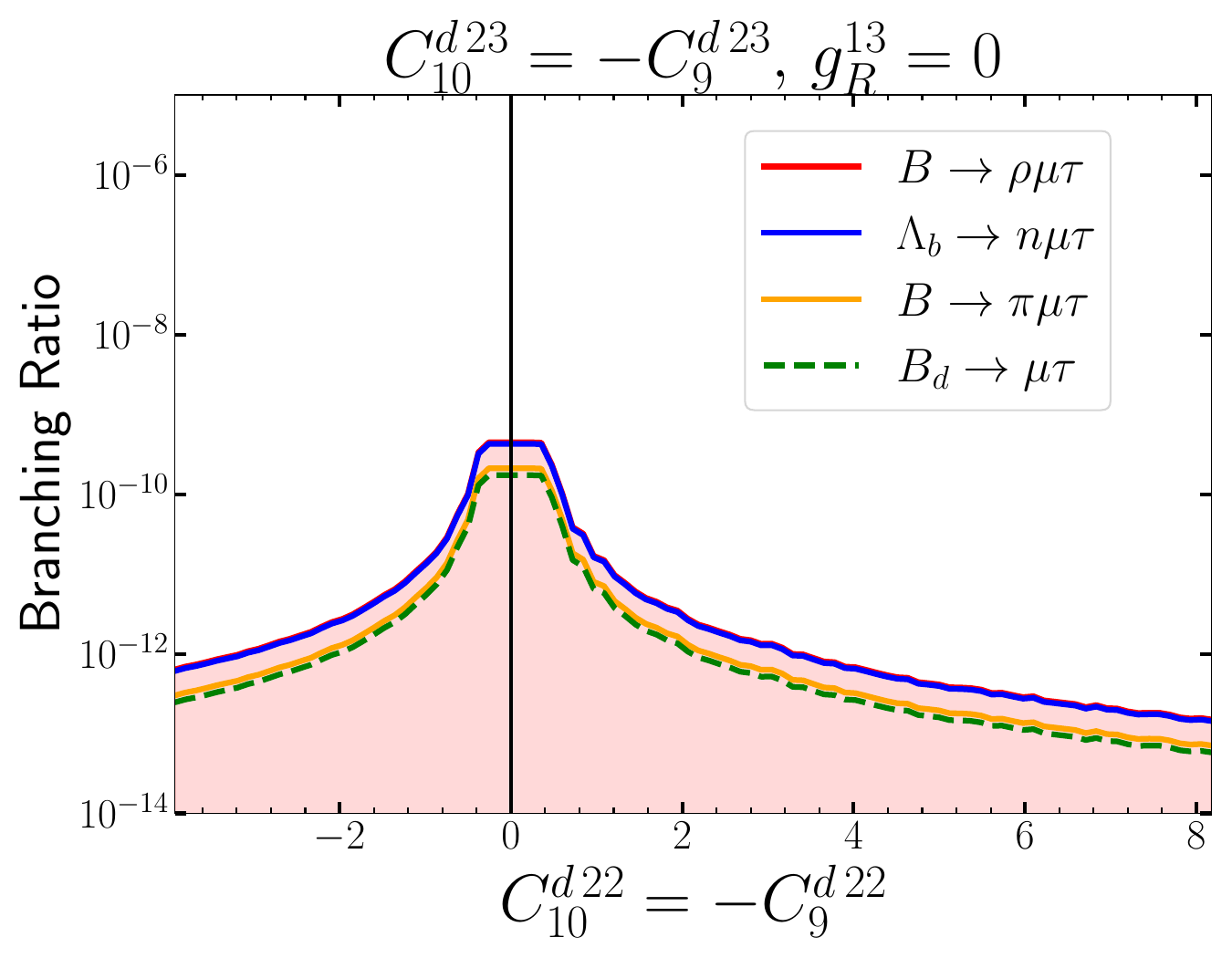}~\includegraphics[width=0.5\linewidth]{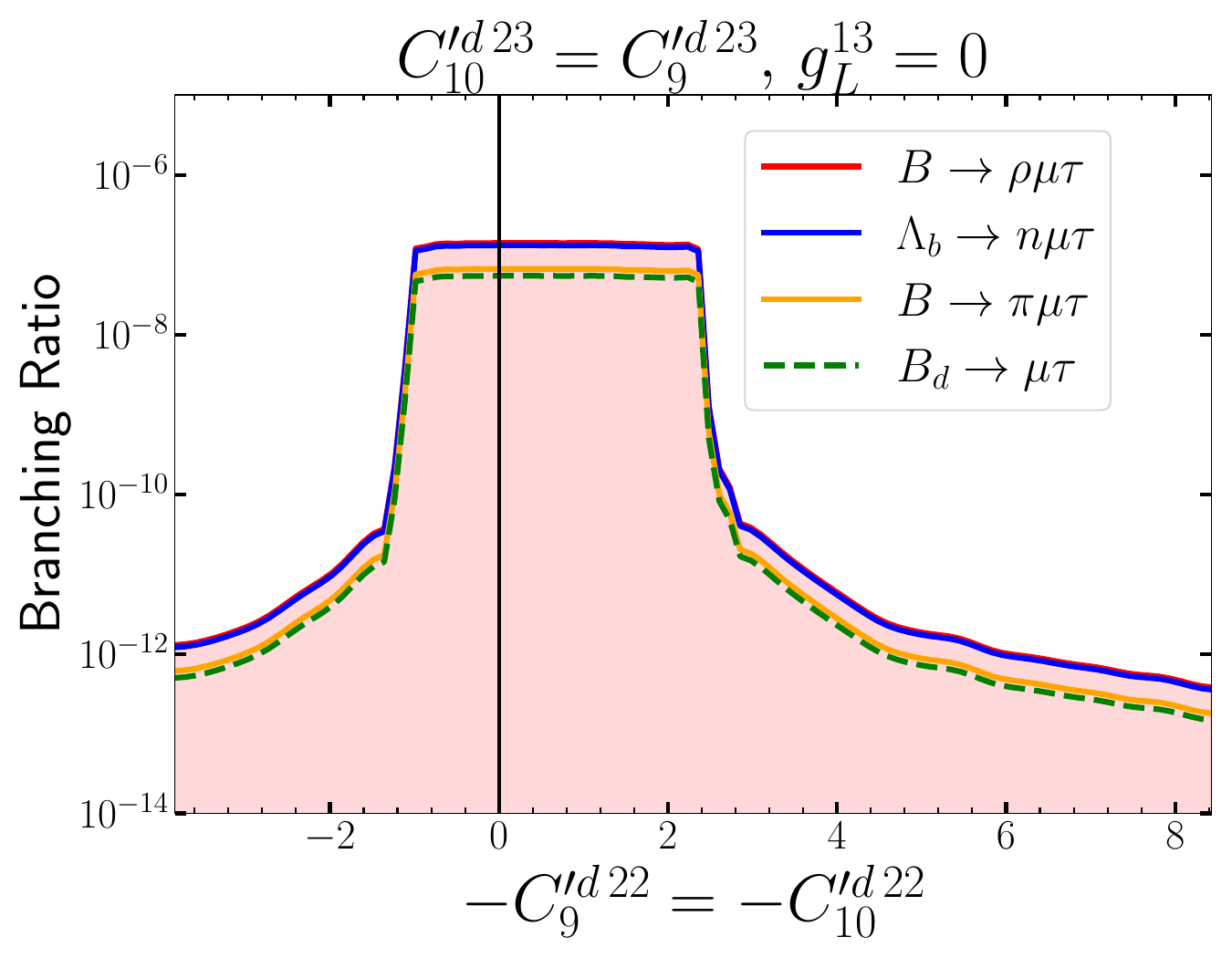}
\caption{\small \sl Upper bounds on the branching fractions for the $b$-hadron LFV decays based on the $b\to s\mu\tau$ (upper row) and $b\to d\mu\tau$ (lower row) transitions, as a function of the Wilson coefficients $C_{9^{(\prime)}}^{q\, 22 }$ and  $C_{{10}^{(\prime)}}^{q\, 22 }$ for the two scenarios defined in the text: (i) Scenario I defined with purely left-handed couplings (left column) and (ii) Scenario II defined with purely right-handed couplings. See text for details.  }
\label{fig:c9c10bs} 
\end{figure}

(i) By combining the measured oscillation frequencies of the $B^0_{s(d)}$--$\overline{B^0}_{s(d)}$ systems, $\Delta m_{B_s}^\mathrm{exp}=17.765(6)~\mathrm{ps}^{-1}$ and $\Delta m_{B_d}^\mathrm{exp}=0.5069(19)~\mathrm{ps}^{-1}$~\cite{ParticleDataGroup:2022pth}, with the expression from Refs.~\cite{Becirevic:2016zri,Aebischer:2020dsw}, one can constrain the non-diagonal couplings to quarks as, 
\begin{align}\label{constraint:bs-bd}
 \dfrac{|g_{L/R}^{bs}|}{m_{Z^\prime}}< {3.4\times 10^{-3}\;\mathrm{TeV}^{-1}} \,,~\qquad\qquad  \dfrac{|g_{L/R}^{bd}|}{m_{Z^\prime}}< 8.5\times 10^{-4} \;\mathrm{TeV}^{-1} \,,
\end{align}

 \noindent where we used the SM predictions, $\Delta m_{B_s}^\mathrm{SM}=18.1(1.1)~\mathrm{ps}^{-1}$ and $\Delta m_{B_d}^\mathrm{SM}=0.54(6)~\mathrm{ps}^{-1}$, based on the lattice QCD estimates for the bag parameters with $N_f=2+1$~\cite{FlavourLatticeAveragingGroupFLAG:2021npn}, and the CKM values $|V_{tb} V_{ts}^\ast|=0.0396(3)$ and $|V_{tb} V_{td}^\ast|=0.0084(3)$, as obtained by using $|V_{cb}|$ from $B\to D \ell \nu$ decays~\cite{FlavourLatticeAveragingGroupFLAG:2021npn,Becirevic:2023aov} and $\gamma$ from Ref.~\cite{UTfit:2022hsi}.~\footnote{See Ref.~\cite{Buras:2022qip} for a detailed discussion of various possible choices for the CKM inputs and their consequences for $\Delta F=1$ and $\Delta F=2$ observables.} Note that the same constraint is relevant to both of our scenarios.

\begin{table}[!t]
\small
\renewcommand{\arraystretch}{1.6}
\centering
\begin{tabular}{|c|cc||c|cc|}
\hline 
Decay  & Scenario I & Scenario II & Decay  & Scenario I & Scenario II\\
\hline\hline
$B_s\to \mu^\pm \tau^\mp$ & $5\times 10^{-8}$& $1\times 10^{-6}$ & $B_d\to \mu^\pm \tau^\mp$ & $2\times 10^{-10}$& $5\times 10^{-8}$\\ 
$B^+\to K^+  \mu^\pm \tau^\mp$ &$6\times10^{-8}$ &$1\times10^{-6}$  & $B^+\to \pi^{+}  \mu^\pm \tau^\mp$ & $2\times 10^{-10}$ &$7\times 10^{-8}$\\
$B^0\to K^{\ast 0}  \mu^\pm \tau^\mp$ & $1\times10^{-7}$&$3\times10^{-6}$& $B^+\to \rho^{+}  \mu^\pm \tau^\mp$  & $5\times 10^{-10}$ & $1\times 10^{-7}$\\
$\Lambda_b\to \Lambda \mu^\pm \tau^\mp$ & $9\times10^{-8}$ & $2\times10^{-6}$& $\Lambda_b\to n  \mu^\pm \tau^\mp$ & $4\times 10^{-10}$ & $1\times 10^{-7}$\\ \hline
\end{tabular}
\caption{\small \sl Maximal branching fractions allowed by the current constraints on the two $Z^\prime$ scenarios that we consider, namely a $Z^\prime$ with left-handed (scenario I) right-handed fermions (scenario II) couplings only. }
\label{tab:final-predictions-Zp} 
\end{table}

 (ii) To constrain the left-handed couplings to leptons, $g_{L}^{ij}$, the leptonic decays $\tau \to \ell \nu \bar{\nu}$ (with $\ell=e,\mu$) are particularly useful as they receive a tree-level contribution mediated by $Z^\prime$. After combining $\mathcal{B}(\tau\to e \nu\bar{\nu})^\mathrm{exp}=17.82(5)\,\%$ and $\mathcal{B}(\tau\to \mu \nu\bar{\nu})^\mathrm{exp}=17.33(5)\,\%$~\cite{ParticleDataGroup:2022pth}, with the expression given in Ref.~\cite{Becirevic:2016zri}, we obtain the $2\sigma$ bounds:
\begin{align}
  \dfrac{|g_{L}^{13}|}{m_{Z^\prime}}< 2\times 10^{-1}~\mathrm{TeV}^{-1} \,,~\qquad\qquad  \dfrac{|g_{L}^{23}|}{m_{Z^\prime}}< 7\times 10^{-1}~\mathrm{TeV}^{-1} \,.
\end{align}

\noindent To constrain the right-handed couplings to leptons, $g_{R}^{ij}$, one can use the experimental bounds $\mathcal{B}(\tau\to e\mu\mu)^\mathrm{exp}<2.7\times 10^{-8}$ and $\mathcal{B}(\tau\to \mu\mu\mu)^\mathrm{exp}<2.1\times 10^{-8}$ (90$\%$ CL)~\cite{ParticleDataGroup:2022pth}, to which $Z^\prime$ contributes at tree level. These bounds and the expressions derived in Ref.~\cite{Crivellin:2015era} yield:
\begin{align}
\text{{Scenario I}}:\qquad\quad  &\dfrac{|g_{L}^{13} g_{L}^{22}|}{m_{Z^\prime}^2}<  1.2\times 10^{-2}~\mathrm{TeV^{-2}} \,,  & \dfrac{|g_{L}^{23} g_{L}^{22}|}{m_{Z^\prime}^2}< 8\times 10^{-3}~\mathrm{TeV^{-2}}  \,,& &\\[0.35em]
\text{{Scenario II}}:\qquad \quad &\dfrac{|g_{R}^{13}g_{R}^{22}|}{m_{Z^\prime}^2}<  1.2\times 10^{-2}~\mathrm{TeV^{-2}} \,,~ &\dfrac{|g_{R}^{23}g_{R}^{22}|}{m_{Z^\prime}^2}< 8\times 10^{-3}~\mathrm{TeV^{-2}} \,.& &
\end{align}

\noindent To limit the diagonal couplings $g_{L/R}^{\mu\mu}$ one can use $\overline{\mathcal{B}}(B_{d(s)}\to\mu\mu)$, together with the bounds from Eq.\eqref{constraint:bs-bd}. 
To that end we use the measured $\overline{\mathcal{B}}(B_{s}\to\mu\mu)^{\mathrm{exp}}=3.35(27)\times 10^{-9}$~\cite{ATLAS:2018cur}, and the upper bound $\overline{\mathcal{B}}(B_{d}\to\mu\mu)^{\mathrm{exp}}<1.9    \times 10^{-10}$ ($95\%$~CL)~\cite{CMS:2022mgd}. From the comparison with the SM predictions $\overline{\mathcal{B}}(B_{s}\to\mu\mu)^{\mathrm{SM}}=3.38(9)\times 10^{-9}$ and $\overline{\mathcal{B}}(B_{d}\to\mu\mu)^{\mathrm{SM}}=9.6(7)\times 10^{-11}$, and using the theory inputs from Refs.~\cite{Beneke:2019slt} except for the CKM matrix elements, for which we use the values mentioned above, 
we obtain: 
\begin{equation}
C_{10}^{d\,22}-C_{10^\prime}^{d\,22}\in (-4,12)\,, \qquad C_{10}^{s\,22}-C_{10^\prime}^{s\,22}\in (-0.3,0.4)\,,
\end{equation}
at 90$\%$ C.L.
These results allow us to constrain the $Z^\prime$ couplings to $\bar{s}b$ and  $\bar{\mu}\mu$ via Eqs.~\eqref{eq:Zp-C10} and \eqref{eq:Zp-C9}.

Now it is a simple matter to combine the low energy constraints (i) and (ii) in both of our scenarios and obtain the bounds for the exclusive $b\to q \ell_i\ell_j$ ($q=s,d$). In Fig.~\ref{fig:c9c10bs} we show the maximally allowed branching fractions for various exclusive modes as a function of $C_{9^{(\prime)}}^{q\,22}=-C_{10^{(\prime)}}^{q\,22}$ and $C_{9^{(\prime)}}^{q\,22}=+C_{10^{(\prime)}}^{q\,22}$, i.e. for the Scenarios I and II. We find that the branching fractions can be as large as $\mathcal{O}(10^{-6})$ and remain consistent with the available low energy constraints. These bounds are summarized in Table~\ref{tab:final-predictions-Zp}. We have also checked that these large branching fractions are consistent with the high-$p_T$ limits derived from the high-energy tails of $pp\to\tau\ell$ (with $\ell=e,\mu$)~\cite{Allwicher:2022gkm}, for which $m_{Z^\prime}\gg 1$~TeV has been assumed so that the corresponding Drell-Yan process can be described by the SMEFT. With that assumption we find that the current low energy bounds on $b\to d\ell_i\ell_j$ and $b\to s\ell_i\ell_j$, collected in Table~\ref{tab:exp}, are more stringent than the ones derived from the high-$p_T$ data~\cite{Angelescu:2020uug,Descotes-Genon:2023pen}.

\section{Conclusion}
\label{sec:conclusion}

In this paper we discussed the exclusive LFV modes based on 
$b\to s\ell_i\ell_j$ and $b\to d\ell_i\ell_j$ decays. Their experimental searches are highly important since their observation would be a clear signal of physics BSM. In contrast to their lepton flavor conserving counterparts the theoretical description of these decays does not suffer from couplings to $c\bar c$ or $u\bar u$  resonances. We used the EFT approach to derive the expressions for differential and/or total decay rates of mesons and baryons. Using the best available theoretical information about the hadronic matrix elements for the generic scenario of physics BSM, we were able to provide the {\it magic numbers} in Tables~\ref{tab:magic-numbers-mesons}--\ref{tab:magic-numbers-baryons} which are phenomenologically useful to either constrain the Wilson coefficients, or to combine different modes in order to distinguish the nature of interaction responsible for the LFV decays. While in the decays of mesons some of the magic numbers are zero and some are not, in the case of baryons all of the magic numbers are non-zero. In full generality, we find that the current experimental bounds on the LFV $B_{d,s}$-meson modes
can be combined to deduce the bounds on the decays of baryons.  In that way we find, for example, $\mathcal{B}(\Lambda_b\to \Lambda \mu\tau)\lesssim 3.6 \times 10^{-5}$.

Another interesting phenomenological feature is that the branching fractions exhibit a distinct hierarchical pattern in the case of LFV being mediated by the (axial-)vector operators, cf. Eq.~\eqref{eq:hierarchy-baryon-2}, which becomes reversed when the (pseudo-)scalar operators are considered instead. In both situations the baryon branching fractions are very close in size to the $B$ decays to a lowest vector meson (and the same pair of leptons).

Finally, we provided the bounds on the LFV modes discussed in this paper in two different models. 
We first considered the scenario in which LFV is mediated by the SM-like Higgs boson. The relevant LFV couplings can be constrained by the current experimental bounds on $\mathcal{B}(h\to \mu\tau)$ and $\mathcal{B}(h\to e\tau)$. We find that the combined experimental information for these two modes to $1\sigma$, provided by ATLAS, is not compatible with the experimental bound on $\mathcal{B}(\mu\to e\gamma)$. We therefore assumed either $e\tau$- or $\mu\tau$-channel to be forbidden, and found that all the LFV decays considered here have tiny branching fractions, well below the experimental reach.

In the second scenario we consider LFV mediated by the exchange of a heavy $Z^\prime$. Using the low energy constraints on the relevant couplings we derived the bounds on $b\to q\mu\tau$ exclusive modes ($q=s$ or $d$), and found that in the case of left-handed couplings to $Z^\prime$, $\mathcal{B}(\Lambda_b\to \Lambda \mu\tau) \lesssim 2\times 10^{-9}$, while for the right-handed couplings to $Z^\prime$ we get $\mathcal{B}(\Lambda_b\to \Lambda \mu\tau) \lesssim 9\times 10^{-7}$, and the bounds become somewhat larger on the corresponding $b\to d$ modes. Clearly, these bounds are more restrictive albeit consistent with Table \ref{tab:indirect-bounds-EFT}, and not too far from the experimental reach of LHCb.

\section*{Acknowledgments}
\label{sec:acknowledgment}

This project has received funding from the European Union’s Horizon Europe research and innovation program under the Marie Sklodowska-Curie Staff Exchange grant agreement No 101086085 – ASYMMETRY and support from the European Union’s Horizon 2020 research and innovation program under the Marie Sklodowska-Curie grant agreement No 860881-HIDDeN.
D.B. thanks IPMU (University of Tokyo) for hospitality where a part of this work has been done.
\newpage
\appendix 
\section{Double differential distribution of $\Gamma(\Lambda_b \to \Lambda \ell_i\ell_j)$\label{app:angular}}

Here we write the expressions for the double differential decay rate of $\Lambda_b (p)\to \Lambda (k) \ell_i^+(p_i)\ell_j^-(p_j)$, with respect to $q^2 = (p_i+p_j)^2=(p -k)^2$, and to $\cos\theta$, where $\theta$ is the angle between the $\ell_i$ and the $z$-axis in the frame $\vec p_i+\vec p_j=0$, with $z$-axis being the direction of flight of $\Lambda$ in the $\vec p=0$ frame. After summing over all spins, we have
\begin{equation}
\frac{d\Gamma(\Lambda_b \to \Lambda  \ell_i^+\ell_j^-)}{dq^2 \, d\cos\theta} = a_\Lambda(q^2) + b_\Lambda(q^2)\cos\theta + c_\Lambda(q^2)\cos^2\theta\,,
\end{equation}
where
\begin{align}
\begin{split}
a_{\Lambda}(q^2) =&\mathcal{N}\Big[ \beta_-^\ell\frac{(m_i+m_j)^2}{q^2}\Big(2|\widetilde{H}^{At}_{--}|^2+2|\widetilde{H}^{At}_{++}|^2+|\widetilde{H}^{V+}_{-+}-\widetilde{H}^{V-}_{-+}|^2+|\widetilde{H}^{V+}_{+-}|^2\Big)\\
&+\beta_-^\ell\Big(2|\widetilde{H}^{V0}_{--}|^2+2|\widetilde{H}^{V0}_{++}|^2+|\widetilde{H}^{V+}_{-+}+\widetilde{H}^{V-}_{-+}|^2\Big)\\
&+\beta_+^\ell\frac{(m_i-m_j)^2}{q^2}\Big(2|\widetilde{H}^{Vt}_{--}|^2+2|\widetilde{H}^{Vt}_{++}|^2+|\widetilde{H}^{A+}_{-+}-\widetilde{H}^{A-}_{-+}|^2+|\widetilde{H}^{A+}_{+-}|^2\Big) \\
&+\beta_+^\ell\Big(2|\widetilde{H}^{A0}_{--}|^2+2|\widetilde{H}^{A0}_{++}|^2+|\widetilde{H}^{A+}_{-+}+\widetilde{H}^{A-}_{-+}|^2\Big) \Big]\,,
\end{split}\\[+3mm]
\begin{split}
b_{\Lambda}(q^2) =&4 \mathcal{N}\sqrt{\lambda_\ell}\, \Big[  \frac{m_j^2-m_i^2}{q^2}{\rm Re}\Big(\widetilde{H}^{At*}_{--}\widetilde{H}^{A0}_{--}+\widetilde{H}^{At*}_{++}\widetilde{H}^{A0}_{++}+\widetilde{H}^{Vt*}_{--}\widetilde{H}^{V0}_{--}+\widetilde{H}^{Vt*}_{++}\widetilde{H}^{V0}_{++}\Big)\\
&\qquad + {\rm Re}\Big(\widetilde{H}^{V-*}_{-+}\widetilde{H}^{A-}_{-+}-\widetilde{H}^{V+*}_{-+}\widetilde{H}^{A+}_{-+}-\widetilde{H}^{V+*}_{+-}\widetilde{H}^{A+}_{+-}\Big)\Big]\,,
\end{split}\\[+3mm]
\begin{split}
c_{\Lambda}(q^2) =&\mathcal{N}\frac{\sqrt{\lambda_\ell}}{q^2}\Big(  |\widetilde{H}^{V+}_{+-}|^2-|\widetilde{H}^{V0}_{++}|^2-2|\widetilde{H}^{V0}_{--}|^2+|\widetilde{H}^{V+}_{-+}-\widetilde{H}^{V-}_{-+}|^2 \\
&+ |\widetilde{H}^{A+}_{+-}|^2-|\widetilde{H}^{A0}_{++}|^2-2|\widetilde{H}^{A0}_{--}|^2+|\widetilde{H}^{A+}_{-+}-\widetilde{H}^{A-}_{-+}|^2\Big) \,,
\end{split}
\end{align}
where, as already defined,  
$\beta^\ell_\pm=q^2-(m_{\ell_i}\pm m_{\ell_j})^2$, and $\lambda_\ell = \beta^\ell_+\beta^\ell_-$.~\footnote{The reader should not be confused that $\lambda_\Lambda$ is here used to label the polarization state of the baryon $\Lambda$ in the final state, while in Eq.\eqref{eq:Lb-semilep} it was a short notation for the triangle function $\lambda^\Lambda\equiv \lambda(\sqrt{q^2}, m_{\Lambda_b},m_{\Lambda})$. 
} The factor 
$\mathcal{N} = \alpha^2 G_F^2 |V_{tb}V_{ts}^\ast|^2/(8192\pi^5 m_{\Lambda_b}^3)$, and the modified $q^2$-dependent helicity amplitudes $\widetilde H_{\lambda_{\Lambda_b} \lambda_{\Lambda}}^{J \lambda } \equiv \widetilde H_{\lambda_{\Lambda_b} \lambda_{\Lambda}}^{J \lambda }(q^2)$ are labeled by $\lambda_{\Lambda_b}$,$\lambda_\Lambda$ and $\lambda$, the respective polarization states of two baryons, $\Lambda_b$, $\Lambda$, and of the intermediate generic vector boson, and $J$ the transition operator, vector or axial current. For $\lambda = t$ or $(\lambda_{\Lambda_b}, \lambda_\Lambda)\in \lbrace{ ++,--\rbrace}$, we have:
\begin{align}
\begin{split}
\widetilde{H}^{Vt}_{++} &= (C_9+C_9')H^{Vt}_{++}+(-C_9+C_9')H^{At}_{++}+\frac{\sqrt{q^2}}{m_{\ell_i}-m_{\ell_j}}\left[(C_S+C_S')H^S_{++}+(C_S-C_S')H^P_{++}\right]\,,\\
\widetilde{H}^{Vt}_{--} &= (C_9+C_9')H^{Vt}_{--}+(-C_9+C_9')H^{At}_{--}+\frac{\sqrt{q^2}}{m_{\ell_i}+m_{\ell_j}}\left[(C_S+C_S')H^S_{--}+(C_S-C_S')H^P_{--}\right]\,,\\
\widetilde{H}^{At}_{++} &= (C_{10}+C_{10}')H^{Vt}_{++}+(-C_{10}+C_{10}')H^{At}_{++}+\frac{\sqrt{q^2}}{m_{\ell_i}+m_{\ell_j}}\left[(C_S+C_S')H^S_{++}+(C_S-C_S')H^P_{++}\right]\,,\\
\widetilde{H}^{At}_{--} &= (C_{10}+C_{10}')H^{Vt}_{--}+(-C_{10}+C_{10}')H^{At}_{--}+\frac{\sqrt{q^2}}{m_{\ell_i}-m_{\ell_j}}\left[(C_S+C_S')H^S_{--}+(C_S-C_S')H^P_{--}\right]\,.
\end{split}
\end{align}
Otherwise, 
\begin{align}
\begin{split}
\widetilde{H}^{V\lambda}_{\lambda_{\Lambda_b}\lambda_\Lambda} &= (C_9+C_9')H^{V\lambda}_{\lambda_{\Lambda_b}\lambda_\Lambda}+(-C_9+C_9')H^{A\lambda}_{\lambda_{\Lambda_b}\lambda_\Lambda}\,,\\[0.4em]
\widetilde{H}^{A\lambda}_{\lambda_{\Lambda_b}\lambda_\Lambda} &= (C_{10}+C_{10}')H^{V\lambda}_{\lambda_{\Lambda_b}\lambda_\Lambda}+(-C_{10}+C_{10}')H^{A\lambda}_{\lambda_{\Lambda_b}\lambda_\Lambda}\,.
\end{split}
\end{align}
The explicit expressions for the true helicity amplitudes are particularly simple with the adopted decomposition of the hadronic matrix elements, cf. Eqs.(\ref{eq:FF-baryon-1}-\ref{eq:FF-baryon-4}):
\begin{align}
H_{++}^{V0} &= H_{--}^{V0} = f_+(q^2)\frac{(m_{\Lambda_b}+m_\Lambda)\sqrt{\beta^\Lambda_-}}{\sqrt{q^2}}\,,\quad\quad
& H_{++}^{V0} &= H_{--}^{Vt} = f_0(q^2)\frac{(m_{\Lambda_b}-m_\Lambda)\sqrt{\beta^\Lambda_+}}{\sqrt{q^2}}\,,\nonumber\\
H_{+-}^{V-} &= H_{-+}^{V+} = -\sqrt{2}f_\perp (q^2) \sqrt{\beta^\Lambda_-}\,,&
H_{++}^{A0} &= -H_{--}^{A0} = g_+(q^2)\frac{(m_{\Lambda_b}-m_\Lambda)\sqrt{\beta^\Lambda_+}}{\sqrt{q^2}}\,,\\
H_{++}^{A0} &= -H_{--}^{At} = g_0(q^2)\frac{(m_{\Lambda_b}+m_\Lambda)\sqrt{\beta^\Lambda_-}}{\sqrt{q^2}}\,, &
H_{+-}^{A-} &= -H_{-+}^{A+} = \sqrt{2}g_\perp (q^2) \sqrt{\beta^\Lambda_-}\,,\nonumber\\
H_{++}^S &= H_{--}^{S} = f_0(q^2)\frac{(m_{\Lambda_b}-m_\Lambda)\sqrt{\beta^\Lambda_+}}{m_b-m_s}\,, &
H_{++}^P &= -H_{--}^{P} = g_0(q^2)\frac{(m_{\Lambda_b}+m_\Lambda)\sqrt{\beta^\Lambda_-}}{m_b+m_s}\,.\nonumber
\end{align}
As in the text, $\beta_{\pm}^\Lambda = (m_{\Lambda_b}\pm m_{\Lambda})^2 - q^2$, and the only remaining ingredients are the quark masses~\cite{FlavourLatticeAveragingGroupFLAG:2021npn}: $m_b^{\overline{\rm MS}}(m_b)=4.20(1)$~GeV, $m_s^{\overline{\rm MS}}(m_b)=76.8(6)$~MeV, $m_{ud}^{\overline{\rm MS}}(m_b)=2.80(3)$~MeV, all given at the common renormalization scale, $\mu=m_b$, after applying the four loop perturbative evolution equation~\cite{Vermaseren:1997fq}.
\newpage 
\section{Alternative set of form factors \label{app:ffs}}

An alternative decomposition of the hadronic matrix elements, often used in the literature, is the following one: 
\begin{align}
\begin{split}
\label{eq:FF-baryon-W}
\langle \Lambda(k,s_\Lambda)| \bar{s}\gamma^\mu b |\Lambda_b(p,s_{\Lambda_b}) \rangle =  &\bar{u}_\Lambda(k,s_\Lambda) \Big[ f_1^V(q^2) \gamma^\mu - f_2^V(q^2) i \sigma^{\mu\nu}\frac{q_\nu}{m_{\Lambda_b}} + f_3^V(q^2) \frac{q_\mu}{m_{\Lambda_b}} \Big] u_{\Lambda_b}(p,s_{\Lambda_b})\,,\\[+3mm]
\langle \Lambda(k,s_\Lambda)| \bar{s}\gamma^\mu \gamma_5 b 
|\Lambda_b(p,s_{\Lambda_b}) \rangle =  &\bar{u}_\Lambda(k,s_\Lambda) \Big[ f_1^A(q^2) \gamma^\mu - f_2^A(q^2) i \sigma^{\mu\nu}\frac{q_\nu}{m_{\Lambda_b}} + f_3^A \frac{q_\mu}{m_{\Lambda_b}} \Big] \gamma_5 u_{\Lambda_b}(p,s_{\Lambda_b})\,.
\end{split}
\end{align}
The above form factors are related to the ones we use in Eqs.(\ref{eq:FF-baryon-1},\ref{eq:FF-baryon-2}) via:
\begin{align}
&f_+(q^2) =f_1^V(q^2)+ \frac{q^2}{m_{\Lambda_b} (m_{\Lambda_b}+m_{\Lambda})}f_2^V(q^2)\,, && g_+(q^2) =f_1^A(q^2)- \frac{q^2}{m_{\Lambda_b} (m_{\Lambda_b}-m_{\Lambda})}f_2^A(q^2)\,,  \nonumber \\[+5mm]
&f_\perp(q^2) =f_1^V(q^2)+ \frac{m_{\Lambda_b}+m_{\Lambda}}{m_{\Lambda_b}} f_2^V(q^2)\,, && g_\perp(q^2) =f_1^A(q^2)- \frac{m_{\Lambda_b}-m_{\Lambda}}{m_{\Lambda_b} }f_2^A(q^2)\,, \nonumber \\[+3mm]
&f_0(q^2) =f_1^V(q^2)+ \frac{q^2}{m_{\Lambda_b} (m_{\Lambda_b}-m_{\Lambda})}f_3^V(q^2)\,, && g_0(q^2) =f_1^A(q^2)- \frac{q^2}{m_{\Lambda_b} (m_{\Lambda_b}+m_{\Lambda})}f_3^A(q^2)\,.  
\end{align}


\end{document}